\documentclass[]{aa}
\usepackage{natbib}
\usepackage[utf8]{inputenc}
\usepackage{graphicx}
\usepackage{subcaption}
\usepackage{txfonts}
\usepackage{color}
\usepackage{amsmath}
\usepackage{comment}

\usepackage{hyperref}
\hypersetup{
	colorlinks   = true, 
	allcolors = blue
}

\bibpunct{(}{)}{;}{a}{}{,} 

\newcommand{\unit}[1]{\,\mathrm{#1}}

\defcitealias{PaperI}{Paper~I}

\begin{document}
	\title{Modeling disks and magnetic outflows around a forming massive star: II. Dynamics of jets from massive protostars}
	\titlerunning{Title in the upper corners of the page}
	\author{André~Oliva \inst{1}
	\and
	Rolf~Kuiper \inst{2,1}
	}

	\institute{Institute for Astronomy and Astrophysics, University of Tübingen, Auf der Morgenstelle 10, D-72076, Tübingen, Germany\\
		\email{andree.oliva@uni-tuebingen.de}  \\
		\and
		Faculty of Physics, University of Duisburg--Essen, Lotharstraße 1, D-47057, Duisburg, Germany\\
		\email{rolf.kuiper@uni-due.de}
	}


	\abstract{
	Forming massive stars launch outflows of magnetic origin, which in fact serve as a marker for finding sites of massive star formation. However, both the theoretical and observational study of the mechanisms that intervene in the formation and propagation of such magnetically-driven outflows has been possible only until recent years.
	}{
	With this work, we aim to study in detail the mechanisms that drive highly collimated outflows from early stages of the formation of a massive star, and how those processes are impacted by the properties of the natal environment of the forming massive star.
	}{
	We perform a series of 31 simulations with the aim of building a unified theoretical picture of these mechanisms, and see how the impact of different environments alter their morphology and momentum output. The magnetohydrodynamical simulations consider also Ohmic dissipation as a nonideal effect, self-gravity, and diffusive radiation transport for thermal absorption and emission by the dust and gas. We start from a collapsing cloud core that is threaded by an initially-uniform magnetic field and which is slowly rotating. We utilize a two-dimensional axisymmetric grid in spherical coordinates.
	}{
	In the simulations, we can clearly distinguish a fast, magneto-centrifugally launched and collimated jet (of speeds $\gtrsim 100\unit{km\,s^{-1}}$), from a wider magnetic tower flow driven by magnetic pressure which broadens in time. We analyze in detail the acceleration of the flow, and its re-collimation by magnetic forces happening at distances of several hundreds of astronomical units. We quantify the impact of magnetic braking in the outflows, which narrows the outflow cavity for the late evolution of the system. We find that despite the non-scalability of self-gravity and the thermodynamics of the medium, our results scale with the mass of the cloud core and can in principle be used with a range of values for such mass. We observe the presence of the same jet-driving mechanisms for a wide range of assumptions on the natal environment of the massive protostar, but with changes to their morphology and mechanical feedback into larger scales over time.
	}{}


	\maketitle

\section{Introduction}

During the formation of massive stars, collimated bipolar outflows are launched, which become a marker of star formation and a source of mechanical feedback into their natal environment. They transport angular momentum into large scales, enhancing accretion through their accompanying disk. There is mounting observational and theoretical evidence, specially in recent years, that highly collimated protostellar outflows are driven by magnetic processes.

The presence of collimated jets around young massive stellar systems is well documented: \cite{Purser2021, Purser2016}, for example, used radio continuum observations to build catalogs of candidates of jets around massive protostars, and \cite{McLeod2018} were able to even detect a jet around a massive young star from an extragalactic source. \cite{Guzman2010} and \cite{Sanna2019} performed radio observations of jets around massive protostars for the star-forming regions IRAS 16562-3959 and G035.02+0.35, respectively. Synchrotron emission has been observed coming from the massive protostellar jets HH 80-81, Cep A HW2 and IRAS 21078+5211 (see \citealt{Carrasco-Gonzalez2010}; \citealt{Carrasco-Gonzalez2021} and \citealt{Moscadelli2021multi}) in regions not close to the protostar (at least a few hundreds of astronomical units away). The origin of this nonthermal emission has been attributed to a re-collimation of the jet due to magnetic forces \citep{Carrasco-Gonzalez2021}. In \cite{Carrasco-Gonzalez2021}, two components of the magnetically-driven outflows are observed: a fast jet (of speeds of at least $500\unit{km\,s^{-1}}$) and a wide-angle wind. Substructures with proper motions have been observed in the outflow ejecta \citep{Obonyo2021}, indicating nonsmoothness in the ejection process.

A new generation of observations that combine data from new or upgraded (sub)millimiter interferometers such as the Northern Extended Millimeter Array (NOEMA), the Atacama Large Millimeter/submillimeter Array (ALMA), the Very Long Baseline Array (VLBI) and the Jansky Very Large Array (JVLA) have enabled an unprecedented resolution of the innermost $\sim 100\unit{au}$ around the forming massive star such that the direct study of the dynamical processes that intervene in the launching \citep{Moscadelli2022} and collimation \citep{Carrasco-Gonzalez2021} of the jet is now possible. However, previous numerical studies of massive protostellar outflows (for example, \citealt{Rosen2020}; \citealt{Machida2020} or \citealt{Mignon-Risse2021II}; a more detailed literature review is offered in Sect. \ref{S: previous}) have not being able to spatially resolve the launching mechanisms of the fast component of the magnetically-driven collimated jets. The study by \cite{Anders2018} showed that a self-consistent launching of a fast magnetic outflow driven by the magneto-centrifugal mechanism was possible.

Thanks to observations of water masers done in \cite{Moscadelli2022}, the innermost $\sim 100\unit{au}$ region where the jet is launched  can now be observed with a resolution of up to $0.05\unit{au}$. In that article, we presented a simulation of a forming massive star surrounded by a disk-jet system that was able to reach those resolutions and reproduced reasonably well the observed jet kinematics. That simulation was part of a larger set of 31 simulations in total, subjected to a deeper analysis of the physical processes and evolution of the disk and the magnetically-driven outflows.

In a previous article (hereafter \citetalias{PaperI}), we focused on the study of the effects of magnetic fields in the dynamics of the disk. We found that the disk can be distinguished into two vertical layers supported by thermal and magnetic pressure, respectively. Additionally, we extensively studied the size of the disk and the impact of magnetic braking under several star-forming environments. In this article, we focus on the dynamics of the magnetically-driven outflows: their launching processes, acceleration, propagation and termination. The use of the full parameter space we introduced in \citetalias{PaperI} enables us to deepen our understanding of how those processes develop, and how the initial conditions for the gravitational collapse influence the resulting outflow, its structure and propagation. 

In section \ref{S: setup}, we summarize the setup and methods we used in \citetalias{PaperI}. Section \ref{S: physics} deals with a dynamical analysis of the launching, acceleration, propagation and termination of the magnetically-driven outflows, focused on the fiducial case of our simulation catalog. In Section \ref{S: jet parameter scan}, we study how the conditions for the onset of gravitational collapse affect the protostellar outflows produced by the forming massive star. In Section \ref{S: previous}, we offer a literature review and a comparison of our results with previous numerical studies. Section \ref{S: summary} contains a summary and the major conclusions of our work. Finally, we examine the numerical convergence of our results in Appendix \ref{S: convergence}.

\section{Methods and parameter space} \label{S: setup}
Here, we summarize the methods and configuration of the simulation catalog already presented in \citetalias{PaperI}. For a more in-depth discussion of the equations solved, and the solution methods used, we point the reader to that reference.

We simulate the formation of a massive star by following the gravitational collapse of a slowly rotating cloud core of $0.1\unit{pc}$ in radius, threaded by an initially-uniform magnetic field, until the formation of a disk-jet system around the massive (proto)star. At the onset of gravitational collapse (initial configuration), the cloud core has a mass $M_C$, distributed along a density profile $\rho \propto r^{\beta_\rho}$ ($r$ being the spherical radius), a rotational-to-gravitational energy ratio $\zeta$, an angular velocity distribution according to $\Omega \propto R^{\beta_\Omega}$ (with $R$ the cylindrical radius), and a uniform magnetic field parallel to the rotation axis whose strength is determined by the mass-to-flux ratio $\bar \mu$ in units of the critical (collapse-preventing) value. We used the \cite{Machida2007} resistivity model (both with the full temperature and density dependence and with a fixed temperature to save computational power and to allow for a direct comparison with the models by \citealt{Anders2018}), and two other cases for comparison: neglecting the resistivity (ideal MHD) and neglecting magnetic fields.

\begin{figure}
	\centering
	\includegraphics[width=0.7\columnwidth]{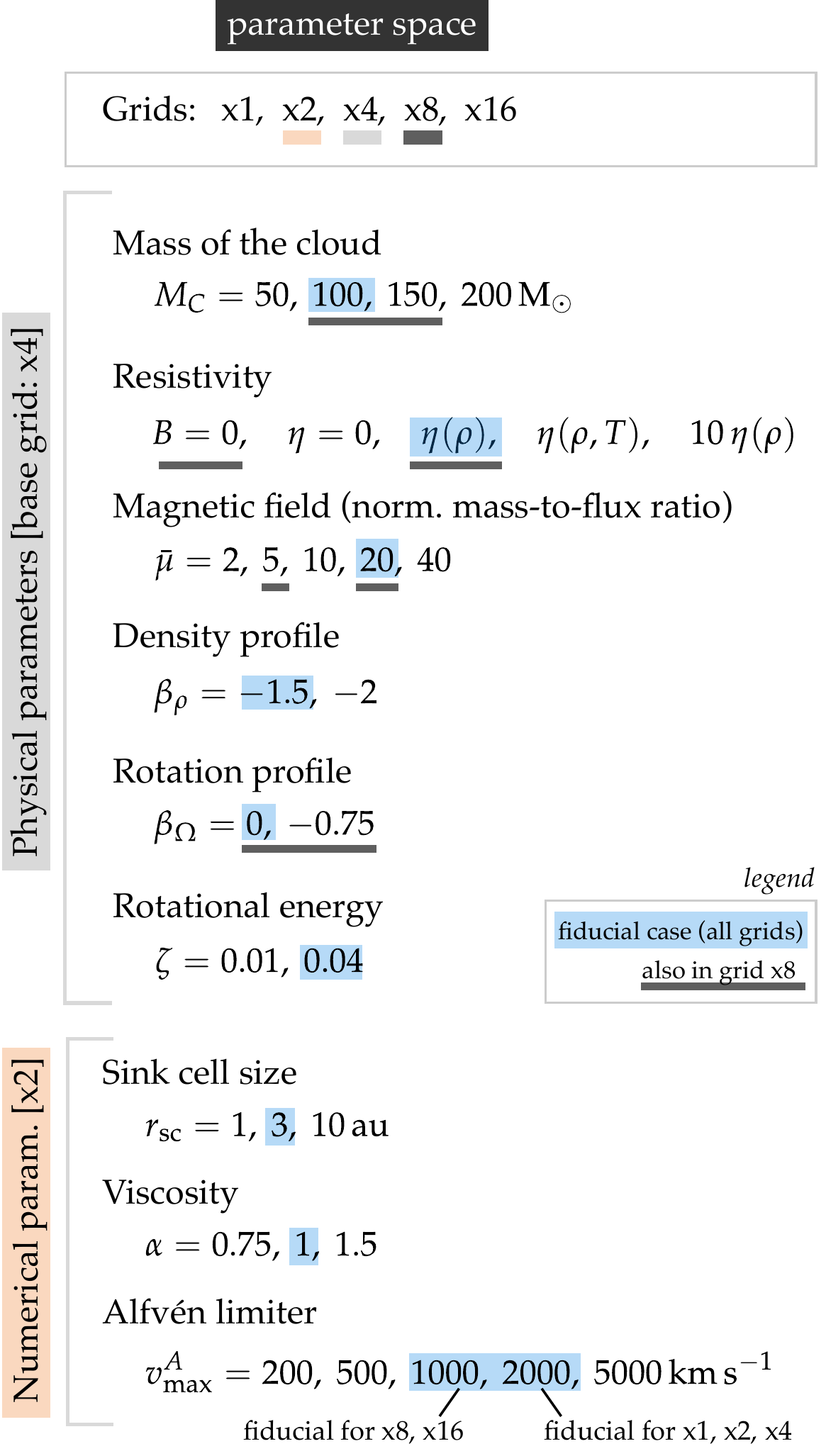}
	\caption{Schematic overview of the parameter space covered by the simulation series.}
	\label{param_space}
\end{figure}

For modeling the weakly ionized gas, we used the methods of magnetohydrodynamics with the code Pluto \citep{Mignone2007}, with additional modules for Ohmic resistivity (with the model by \citealt{Machida2007}), self-gravity \citep{Kuiper2010circ}, and the radiation transport of thermal emission from the dust and gas \citep{Kuiper2020} with the gray flux-limited diffusion approximation. We assume axisymmetry and equatorial symmetry in the simulations. The inner boundary of the simulation domain, which we refer as the sink cell, models the (proto)star. Matter that flows into the sink cell is considered as accreted by the protostar and cannot escape. The magnetic field lines are prolonged across the sink cell with a zero-gradient boundary condition. No flows are artificially injected into the cloud from the protostar; the outflow launching occurs self-consistently within the computational domain. The accretion disk is massive and is expected to form spiral arms and fragment \citep{Oliva2020}. In order to model the gravitational torque from the spiral arms in the azimuthal direction, we use shear $\alpha$-viscosity, following the setup and results of \cite{Kuiper2011}. Inside the outflow cavity, the Alfvén speed increases significantly due to the low densities and strong magnetic fields. In order to avoid extremely small numerical time steps, a limit to the Alfvén speed was used (a detailed discussion of its effects on the outflows is offered in Appendix \ref{S: Alfven}). 

Convergence of the results was studied in the parameter space by using five different grids of increasing resolution, as well as several sizes for the inner boundary (which we refer as the sink cell, and which models the forming protostar).   Grid x16, the highest resolution one in our series, consists of 896 logarithmically spaced cells in the radial direction, and 160 linearly spaced cells along the polar angle. This translates to a minimum grid cell size is $0.03\unit{au}$ at $3\unit{au}$ from the forming massive star and $10\unit{au}$ at $r=1000\unit{au}$. The rest of the grid sizes in the series are obtained by dividing by two the number of cells of each direction. Figure \ref{param_space} presents a schematic summary of the parameter space considered in this study. The values of each parameter corresponding to the fiducial case are highlighted in blue. To simplify nomenclature, we use the following convention: modifications of the fiducial values are written followed by the name of the corresponding grid. For example, $\bar \mu = 5$~[x8] means a cloud of a mass-to-flux ratio of 5, with the rest of the parameters as in the fiducial case run on the grid x8.

\section{Physical processes in the outflows} \label{S: physics}
 
 This section is focused on a detailed examination of the outflows produced in the fiducial simulation of our series. We compute several force terms from the magnetohydrodynamical equations and compare their relative strength with the aim of understanding the physical processes that intervene in the launching, acceleration, propagation and termination of the different components of the magnetically-driven massive protostellar outflows.
 
\subsection{Magneto-centrifugal launching} \label{S: mc}
\begin{figure}
	\centering
	\includegraphics[width=\columnwidth]{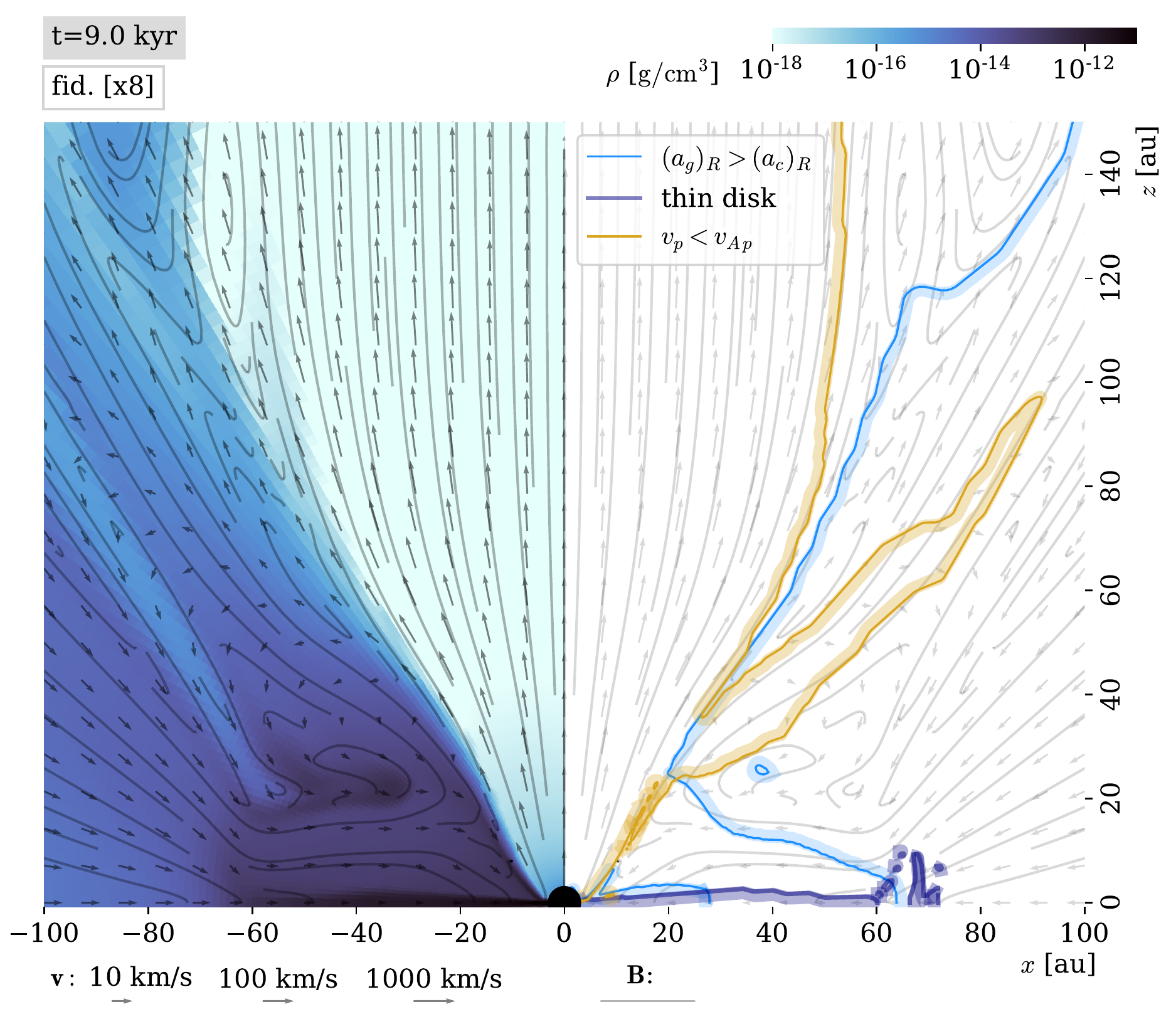}
	\caption{Magneto-centrifugal mechanism for the fiducial case in x8}
	\label{x8_mc}
\end{figure}

The low-density, bipolar jet cavity starts to appear close to the protostar shortly after the formation of the disk at around $t\sim 5\unit{kyr}$, in a process described in more detail in Sect. \ref{S: bow}, which results in an initial bow shock being thrust out. Once the high-speed component of the outflow has established, (and similarly to \citealt{Anders2018}) we find evidence for the magneto-centrifugal mechanism described in \cite{BlandfordPayne1982} inside of the cavity, although with some key geometrical differences which we explain below.

A brief note on the visualizations that make use of directed contours throughout the article: an understanding of the launching mechanisms for outflows requires the simultaneous visualization of several intervening forces, morphological features, and other quantities that carry dynamical information. With this end, and taking advantage of the axial symmetry of our models, we present here a system of plots containing contour lines which aided the interpretation of our results and the drawing of our conclusions. The first instance of such visualizations is given in the right panel of Fig. \ref{x8_mc}. Each solid contour line is followed by an adjacent shadow of the same color, which indicates the direction on which the inequality listed in the legend holds true. Throughout the article, we refer to both the contour and its corresponding shadow as a \emph{directed contour} \footnote{Directed contours are computed with respect to the boundaries of the grid cells, making them appear unnecessarily complex in high-resolution simulations. A smoothing filter has been applied to all the directed contours in this article in order to improve clarity in the explanations. As a caveat, this smoothing process causes some parts of certain contours to appear open.}.

First, we perform a dynamical examination of the launching mechanism of the fast jet. Figure \ref{x8_mc} displays a typical scenario during the magneto-centrifugal phase of the simulation. The left panel provides a joint view of the gas mass density (background color map), poloidal velocity (as a vector field with arrows) and magnetic field (as field lines in solid color), and the right panel presents the directed contours that separate the fulfillment of conditions on two quantities of dynamical interest. The blue directed contour indicates the equilibrium between the (co-rotating) centrifugal acceleration
\begin{equation}
	(a_g)_R = \frac{v_\phi^2}{R},
\end{equation}
and the cylindrical radial component of the gravitational acceleration
\begin{equation}
	(a_c)_R = \frac{GM(r)}{r^2} \sin\theta,
\end{equation}
where $M(r)$ is the mass enclosed in a spherical radius $r$ (including the mass of the protostar), and which constitutes an approximation to the use of the full gravitational potential of stellar gravity and gas self-gravity. In the shadowed blue region, the gravitational force dominates over the centrifugal force along the cylindrical radial direction. A disadvantage of the use of contours to visualize a ratio of forces is the poor handling of regions where equilibrium holds, as opposed to regions with clear dominance of a force over another. This is the case in the thick and thin layers of the disk, for which the contour is no longer adequate. The equilibrium of forces in the disk was studied in detail in \citetalias{PaperI}: in a nutshell, both the thick and thin disks are roughly Keplerian, with thermal and magnetic pressure causing small local regions of sub- and super-Keplerianity, but with the material always being transported inwards.

The yellow directed contour delimits the Alfvén surface, that is the surface where the speed of the plasma is lower than the Alfvén speed
\begin{equation}
	v_A = \frac{|\vec B|}{\sqrt{4\pi \rho}},
\end{equation}
where the magnetic field strength $|\vec B|$ is expressed in Gauss. For the purpose of our analysis, we use the co-rotating frame of reference, which means that we take only the magnitudes of the poloidal components of the velocity and magnetic field\footnote{Given that we assume axisymmetry and the simulations were performed on a grid in spherical coordinates, we decompose a given vector $\vec A$ into two perpendicular vector components: toroidal $A_\phi \vec e_\phi$, and poloidal $\vec A_p = A_r \vec e_r + A_\theta \vec e_\theta$, with $\vec e_i$ denoting unit vectors.}. Outside of the Alfvén surface, perturbations to the originally vertical magnetic field (in this case, caused by the gravitational collapse) occur faster than the restoring force of magnetic tension, resulting in the typical hour-glass morphology of gravitational collapse of a gaseous cloud in rotation \citep[for an observational example, see][]{Beuther2020}. The contrary happens inside the Alfvén surface: the gas flow is forced to follow the magnetic field lines thanks to the restoring magnetic forces. The sub-Alfvénic regime is reached because of the strong magnetic fields close to the protostar (caused by the gravitational drag of the magnetic field) and the low density in the cavity.

The simulations presented include magnetic diffusion in the form of Ohmic resistivity, which acts mostly in the thin and thick disks (see \citetalias{PaperI}) but its action is severely limited in the cavity due to its low density (and therefore, low resistivity). This implies that in the cavity, the plasma is essentially flux-frozen. This can be readily seen while comparing the directions of the velocity field and magnetic field lines in the cavity as presented in Fig. \ref{x8_mc}. We note, however, that while the plasma is flux-frozen in the simulations, this might not be fully the case in reality because other forms of magnetic diffusion (ambipolar diffusion and the Hall effect) are not considered in this study.

We can summarize the magneto-centrifugal mechanism as seen in our results in the following way. Consider a parcel of gas that has fallen in from large scales onto the thick disk. It is then transported through the accretion disk until it reaches the cavity wall. If the parcel enters the cavity, it experiences $(a_c)_R > (a_g)_R$, and the parcel would tend to move cylindrical-radially outwards in absence of magnetic fields. However, the parcel enters simultaneously the sub-Alfvénic regime, which means that the flow must follow the magnetic field lines resulting in the parcel being launched radially outwards. The acceleration of the parcel inside the cavity is detailed in Sect. \ref{S: acc}.

The model by \cite{BlandfordPayne1982} was proposed in the context of accretion disks surrounding black holes, i.e., an infalling envelope was not included, which is necessary in the study of the early stages of protostellar jets. Even though we find general agreement in the launching mechanism of the high-velocity jet, we observe that the jet launching area is not located on top of the full accretion disk. The ram pressure from the infalling envelope, and the effect of magnetic pressure on the disk (the existence of a thick layer in the disk) constrain the formation of the cavity to a conic region in the innermost scales, which means that the launching surface of the jet is located on the portion of the cavity wall that is in contact with the incoming material from the thick accretion disk. The densities found in the envelope and the thick disk impede the flow to become sub-Alfvénic except in the cavity, close to the launching area. The Alfvén surface does not occupy the full cavity, but it is restricted to a narrow region close to the rotation axis and reaches sizes of $\sim 1000\unit{au}$, after which the magnetic field becomes too weak. Beyond the Alfvén surface, the flow is collimated in a process described in Sect. \ref{S: jc}.

\subsection{Acceleration of material in the jet}\label{S: acc}
\begin{figure}
	\centering
	\includegraphics[width=\columnwidth]{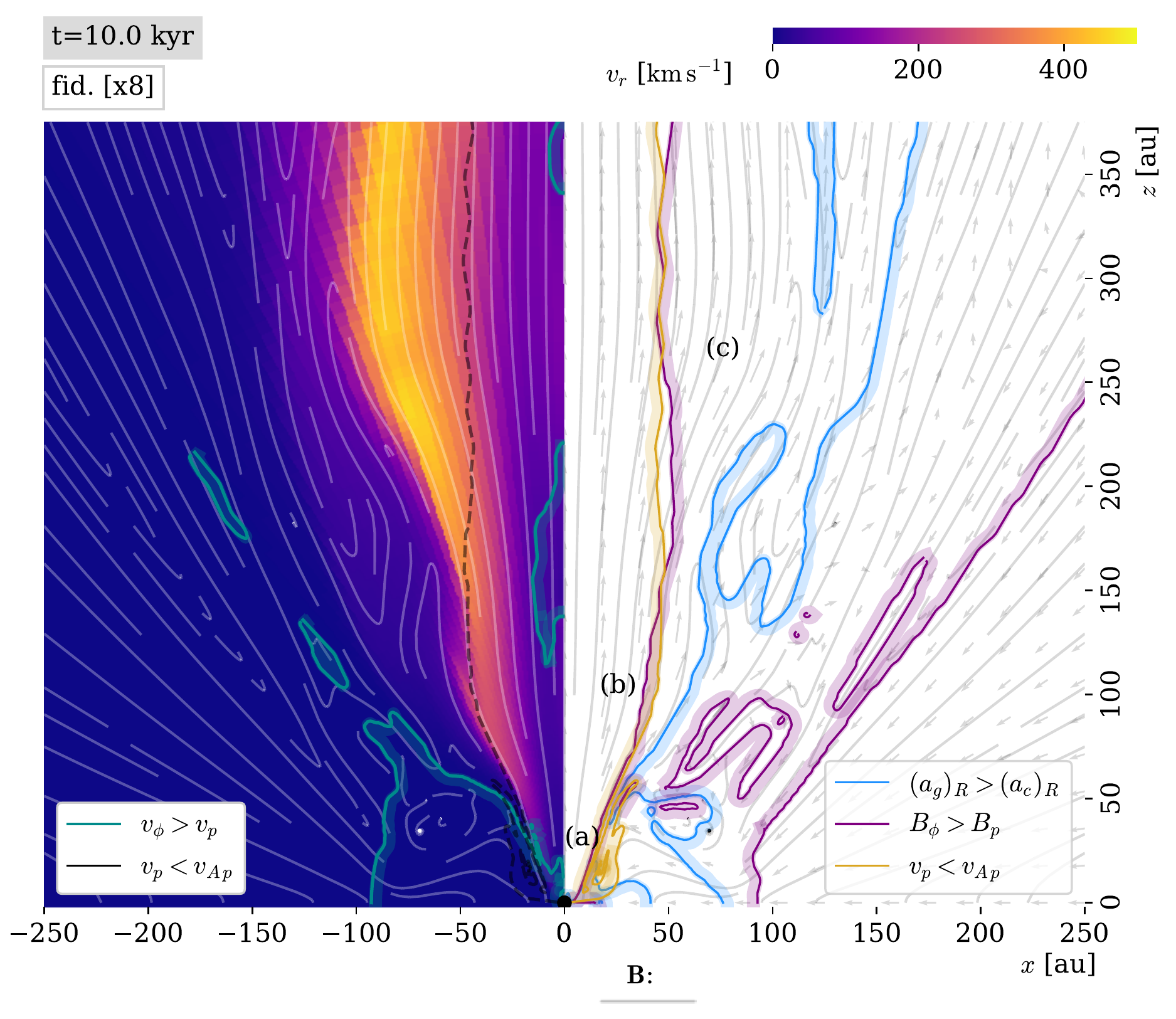}
	\caption{Acceleration of material along the magnetic field lines}
	\label{jet_acceleration_x8}
\end{figure}

In this section, we continue the discussion of the trajectory of a parcel of plasma as it accelerates throughout the cavity and is expelled into larger scales. The right panel of Fig. \ref{jet_acceleration_x8} provides a view of the gravito-centrifugal equilibrium and the Alfvén surface as discussed above, but also presents a directed contour (in purple) that delimits the region where the magnetic field lines are mostly wound by rotation. In the launching area (marked as \emph{a}), the magnetic field lines are mostly poloidal, and a (sub-Alfvénic) parcel of plasma is then forced to follow a mostly radial trajectory until it reaches the Alfvén surface.

To see this explicitly, we take the energy equation of the plasma that, ignoring radiation transport and considering a stationary state, reduces to
\begin{equation} \label{e:eneq-1}
\vec \nabla \cdot \left[ (E^g+E^K+E^\text{th} +  P)\vec v + c \vec{\mathcal E} \times \vec B \right] =0,
\end{equation}
with the first terms inside the parenthesis being the gravitational, kinetic and thermal energy densities, and $\vec {\mathcal E}$ being the electric field. The presence of the gravitational energy density as stated is possible because it does not change significantly over time during the acceleration process. Given that in the cavity we can ignore Ohmic dissipation, $c \vec{\mathcal E}  = -\vec v \times \vec B$. As discussed, the sub-Alfvénicity of the flow close to the launching area causes $\vec B \parallel \vec v$, which implies $c \vec{\mathcal E} \times \vec B = -(\vec v \times \vec B) \times \vec B = \vec 0$ and allows us to conclude from Eq. \ref{e:eneq-1} that
\begin{equation}\label{e:eneq-2}
E^g+E^K+E^\text{th} +  P = \text{constant}
\end{equation}
along the flow in the cavity. This expression enables us to study the process of acceleration of material in the jet, illustrated in the left panel of Fig. \ref{jet_acceleration_x8}. The background map represents the radial velocity (for which the color scale ignores the negative values), and the directed contour highlights the region where the azimuthal component of the velocity is higher in magnitude than its poloidal value. Outside of the disk and in the outer parts of the cavity, the velocity turns from mostly azimuthal to mostly poloidal. As the parcel is forced to follow the magnetic field lines from the position marked as \emph{a} to \emph{b} in Fig. \ref{jet_acceleration_x8}, it travels to larger $r$, where the gravitational energy is lower. Then, according to Eq. \ref{e:eneq-2}, the kinetic energy of the parcel has to increase, given our result from \citetalias{PaperI} that the contribution of the thermal energy and pressure are negligible. The parcel accelerates $v_r \sim 0$ to $\sim 400\unit{km/s}$ along the spherical-radial direction. As a consequence of this process, the highest velocities of the jet at small scales are not found close to the rotation axis.

Once the parcel of plasma leaves the Alfvén surface (mark \emph{c} in Fig. \ref{jet_acceleration_x8}), it enters the region where the magnetic field lines are mostly wound. The trajectory of the flow becomes helical as the flow partially follows the magnetic field lines, but an additional upward acceleration is experienced by the parcel because of magnetic pressure (more details of this process are offered in Sect. \ref{S: tf}). The jet now propagates away from the protostar and into larger scales of the cloud core.

A more thorough examination of the jet speeds across the whole simulation series reveals that speeds of at least $100\unit{km\,s^{-1}}$, and over $1000\unit{km\,s^{-1}}$ are reached in certain regions. However, the maximum speed reached by the jet is not a fully converged value for the reasons discussed in Appendix \ref{S: convergence}.

The most direct observational evidence of the magneto-centrifugal mechanism and the acceleration of the jet as described here is provided by the recent observations of the most massive young stellar object ($M_\star = 5.6\pm 2.0 M_\odot$) in the star-forming region IRAS 21078+5211 \citep{Moscadelli2022, Moscadelli2021multi}. In \cite{Moscadelli2022}, 22 GHz water maser observations (which trace internal shocks of the gas), were able to uncover the helical trajectories of the plasma as it is magneto-centrifugally ejected. In that article, we used the simulation corresponding to the fiducial case with grid x16 to produce a snapshot comparison when the protostellar mass is similar to the observed value, and found that the geometry and kinematics of the transition region where the flow abandons the Alfvén surface and becomes helical coincide very well with the observed masers. Furthermore, the high velocities of the water masers observed in IRAS 21078+5211 ($\sim 200\unit{km\,s^{-1}}$) are fully compatible with the values found in the numerical simulations.

\subsection{Cavity wall ejections} \label{S: we}
\begin{figure}
	\centering
	\includegraphics[width=\columnwidth]{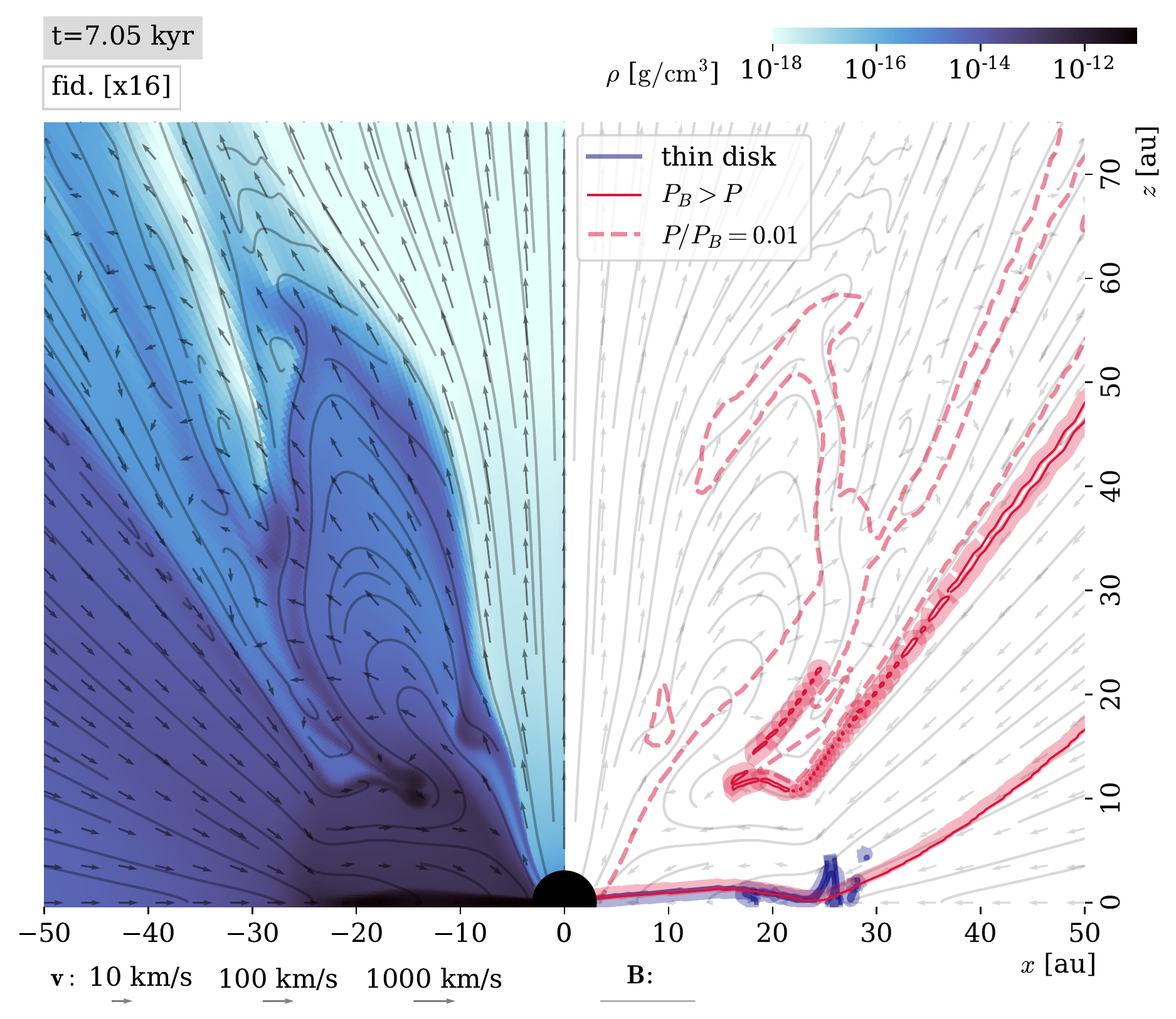}
	\caption{Cavity wall ejection}
	\label{x16_we}
\end{figure}

The cavity wall contributes to both inflows and outflows episodically. When a section from the cavity wall enters the region where centrifugal forces dominate over gravity, the section starts to move outwards, forming a toroidal prominence, as illustrated in Fig. \ref{x16_we}. This prominence grows quickly and develops a neck, thanks to the material infalling from larger scales that provides an inward flow that narrows the neck. The densities present on the cavity wall are sufficient for magnetic diffusion to take place, allowing magnetic field lines to reconnect. This in turn allows the protuberance to decouple from the rest of the cavity wall and it is ejected through the cavity thanks to the magneto-centrifugal mechanism.

The solid pink directed contour in the right panel of Fig. \ref{x16_we} shows the regions where magnetic pressure dominates over thermal pressure, that is, it delimits the region where
\begin{equation}
\beta_p \equiv \frac{P}{P_B} = \frac{E^\text{th}}{E^B} < 1.
\end{equation}
The contour confirms that the thin disk is thermally supported while the thick disk is magnetically supported in the vertical direction, as it was concluded from \citetalias{PaperI}. The densest parts of the cavity wall become thermally dominated, as magnetic diffusion becomes high enough in those regions and then magnetic energy is dissipated into thermal energy. We also display the contour for $\beta_p = 0.01$, which temporarily includes the protuberance and shows that some magnetic energy is dissipated into thermal energy as the magnetic field lines are reconnected. After the ejection event takes place, the magnetic field lines are quickly reconfigured in the cavity wall in a similar way to Fig. \ref{x8_mc}. We remark, however, that the ejection mechanism for the cavity wall ejections is the magneto-centrifugal mechanism and not the energy released by magnetic diffusion during reconnection. The reconnection of magnetic field lines provides a way for the mass to be released into the cavity.

The cavity wall ejections provide not only a mechanism for episodic ejection, but also a mechanism for episodic accretion, as the infall from the envelope onto the protostar is temporarily halted while the ejection event takes place. The observed timescale for the formation and ejection of the protuberance is $\sim 10\unit{yr}$, with the reconnection event happening in $\lesssim 1\unit{yr}$, although those values increase with the elapsed time as the disk grows in size.

We remind the reader that the simulations are performed under the assumption of axisymmetry, and therefore, a three-dimensional study that resolves similar scales would be required to know the role that a turbulent flow would have on ejected material from the cavity wall. For example, turbulent reconnection might make cavity wall ejections more frequent, asymmetric and smaller in size, instead of episodic and large in size, as we find. Additionally, the inclusion of ambipolar diffusion and the Hall effect may affect the reconnection process or not allow the formation of the prominence at all.

\subsection{Magnetic tower flow} \label{S: tf}
\begin{figure}
	\centering
	\includegraphics[width=\columnwidth]{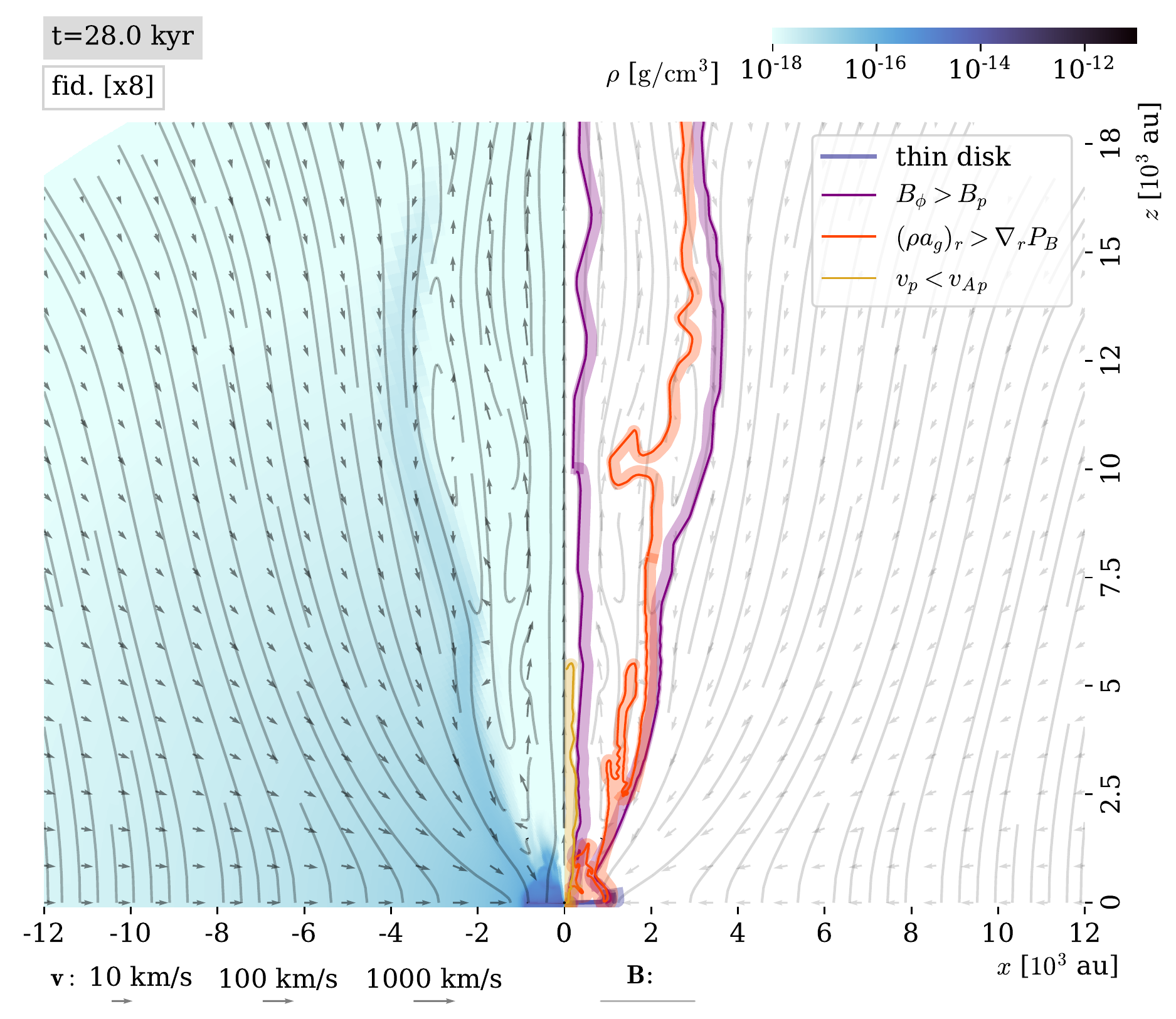}
	\caption{Magnetic tower flow}
	\label{x8_tf}
\end{figure}

We observe a different kind of magnetically-driven outflow which we refer to as the magnetic tower flow, and corresponds to the outflow mechanism described in \cite{Lynden-Bell2003}. As time progresses, the dragging of magnetic field lines by rotation increase magnetic pressure until its gradient is eventually larger than the local gravitational force. At large scales, outside of the Alfvén surface, this force drives the magnetic tower flow.

An example of the tower flow is given in Fig. \ref{x8_tf}, taken from late stages ($t = 28\unit{kyr} = 0.59 t_\text{ff}$) and large scales (the cavity reaches the radius of the cloud). Again, the purple directed contour delimits the region where the magnetic field lines are wound by the rotational drag (i.e., where $|B_\phi| > |B_p|$). We observe that the magnetic field lines are mostly toroidal in most parts of the cavity at large scales, with the exception of the regions close to the rotation axis and also inside of the Alfvén surface, where the magnetic field is mostly poloidal.

The $z$-component of the momentum equation, considering a steady-state flow, reads
\begin{equation}\label{e:mhd-z}
 \nabla \cdot \left(\rho v_z \vec v  - \tfrac{1}{4\pi} {B_z \vec B}\right) + {\partial_z P_t} = \rho a_{g\,z},
\end{equation}
where the divergence $\nabla \cdot \vec f$ of a each vector field $\vec f(R,z)$ is $\nabla \cdot \vec f = (1/R)\partial_R(Rf_R) + \partial_z f_z$. Given that the magnetic field lines are strongly toroidal in the tower flow, we can neglect the contribution from magnetic tension, which only involves the poloidal components. The low thermal pressures at large scales and in the cavity allow us to consider only the contribution of the magnetic pressure to the total pressure gradient. With these approximations, Eq. \ref{e:mhd-z} becomes
\begin{equation}
 \nabla \cdot (\rho v_z \vec v) =  \rho a_{g\,z} - \tfrac{1}{8\pi}\partial_z B^2 
\end{equation}
The $z$-component of the gravitational force is negative in the upper hemisphere. In general, the magnetic field strength increases towards the center of the cloud because it is being dragged there by gravity, and so, the $z$-component of the pressure gradient is generally negative in the same region. Therefore an outward momentum flow is obtained if
\begin{equation}\label{e:tf-cond}
\left|\tfrac{1}{8\pi}\partial_z B^2\right| > |\rho a_{g\,z}|.
\end{equation}

The condition of Eq. \ref{e:tf-cond} is calculated and shown in Fig. \ref{x8_tf} as the non-shadowed part of the orange directed contour. That directed contour was computed with the spherical-radial components of both quantities for simplicity, given that $z\sim r$ for $\theta \to 0$. Indeed, in the tower flow, the magnetic pressure gradient is larger than gravity, and since $B$ is toroidal in the same region, we conclude that the tower flow is driven by the winding of magnetic field lines by rotation.

The speeds obtained in the tower flow are slower than the magneto-centrifugally driven jet. Typical values are of the order of $10 \unit{km\,s^{-1}}$. The morphology and velocity structure of the tower flow can vary significantly with the initial conditions chosen for the cloud core. This is explored in more detail in Sect. \ref{S: jet parameter scan}.

\subsection{Magnetic broadening of the outflow cavity}
\begin{figure}
	\includegraphics[width=\columnwidth]{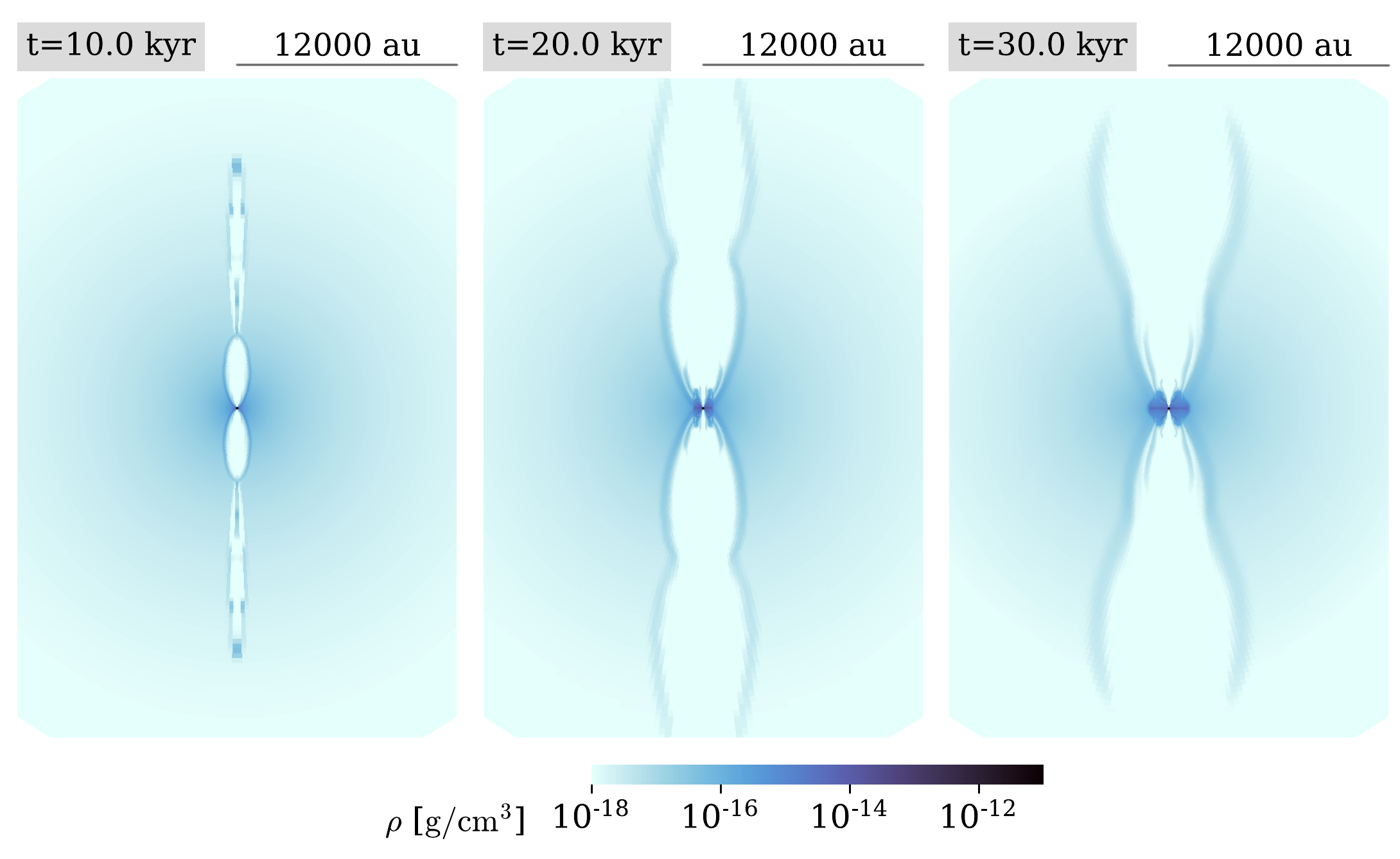}
	\caption{Magnetic broadening of the cavity due to the tower flow.}
	\label{towerflow-broadening}
\end{figure}

We observe (Fig. \ref{towerflow-broadening}) that the tower flow increases its thickness over time. As the accretion disk grows in size with time (cf. \citealt{PaperI}), so does the region of the cloud where the magnetic field lines are wound by rotation. Magnetic pressure keeps increasing in the cavity, strengthening the tower flow and broadening it as a consequence. Additionally, the slow depletion of the envelope means that the ram pressure from large scales in the cloud core decreases over time, allowing further growth of the cavity. The broadening of the tower flow offers an additional component for the earliest outflow broadening (before the driving of HII regions) around early O and B-type stars (see \citealt{Beuther2005}). According to our results, very collimated outflows are indicative of the earliest stages of star formation (fiducial case $t\sim 10 \unit{kyr}$ or $0.2 t_\mathrm{ff}$). A first broadening comes as a result of the broadening of the tower flow ($t\gtrsim 20\unit{kyr}$), and we expect a second broadening due to the growth of the HII region ($t\gtrsim 100\unit{kyr}$, see \citealt{KuiperHosokawa2018}) before other forms of radiative feedback (e.g. stellar winds and radiation pressure) take over.

\subsection{Jet re-collimation} \label{S: jc}
\begin{figure}
	\centering
	\includegraphics[width=\columnwidth]{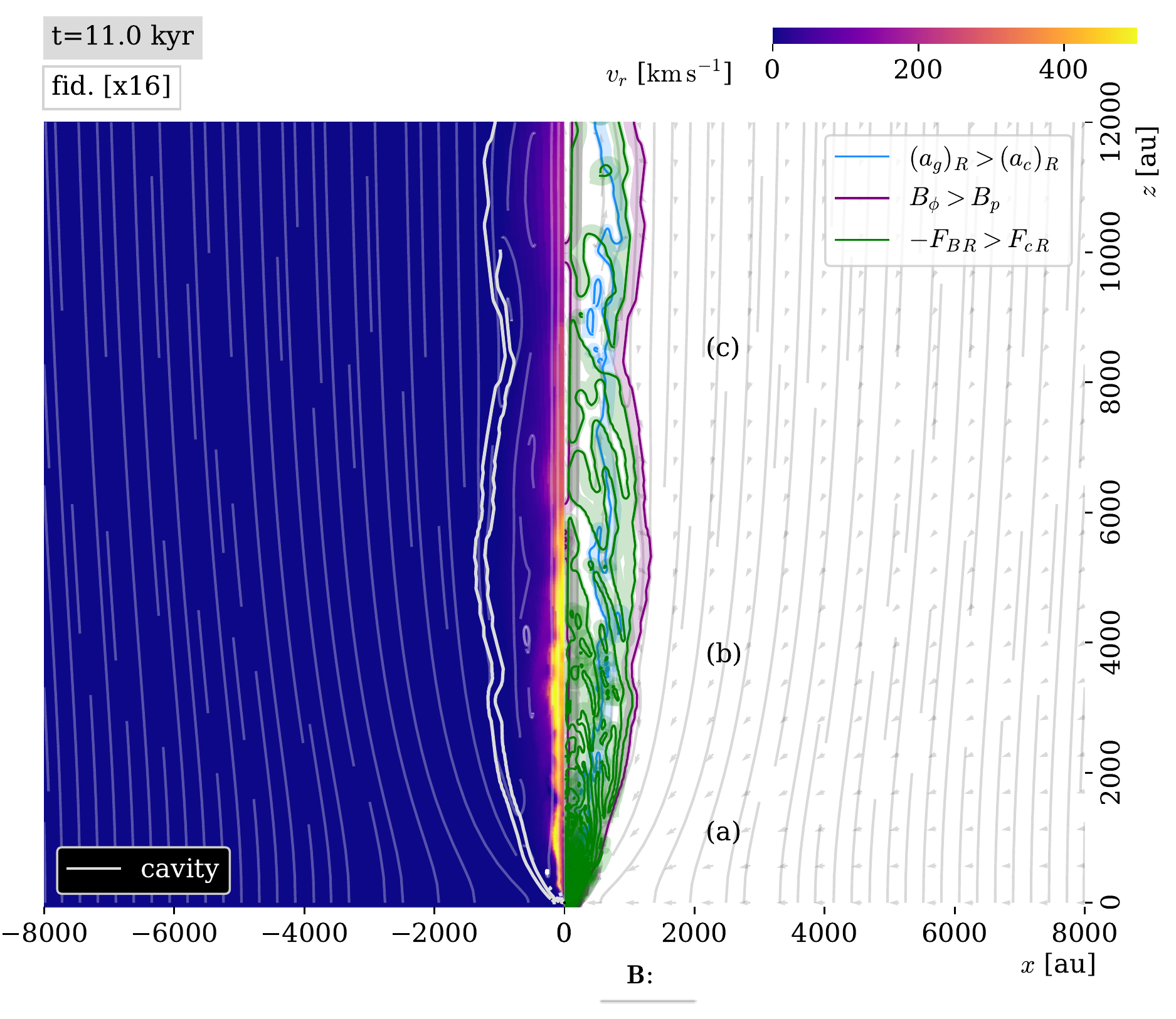}
	\caption{Re-collimation of the jet. The color scale for the radial velocity is saturated at $v_r = 500\unit{km\,s^{-1}}$. For a fuller picture of the speeds reached over time, we refer the reader to Fig. \ref{morph-x8}.}
	\label{x16_jc}
\end{figure}

At the launching region, the jet is accelerated along the spherical-radial direction, which means that the highest speeds are not reached along the rotation axis. However, at scales of a few thousand astronomical units along the propagation of the jet, and fully outside of the Alfvén surface, we find evidence that magnetic hoop stress and the ram pressure from the infalling envelope exert forces that re-collimate the flow and yield high radial velocities at the rotation axis. In Fig. \ref{x16_jc}, we observe those re-collimation points marked as \emph{a}, \emph{b} and \emph{c}, where the cavity appears narrower. The left panel of Fig. \ref{x16_jc} presents the positive radial velocity, which peaks particularly strongly around the re-collimation points \emph{a} and \emph{b}. The right-hand panel of the figure provides a visualization of the forces involved in the re-collimation. Similarly to the previous analysis, the region where the magnetic field lines are wound is shown in the shadowed area of the purple directed contour. Additionally, the green directed contour displays the regions where the cylindrical-radial component of the Lorentz force density
\begin{equation} \label{eq: FBR}
F_{BR} = (\vec J \times \vec B)_R /c=  \tfrac{1}{4\pi} [ (\nabla \times \vec B ) \times \vec B]_R
\end{equation}
is negative (i.e., directed inwards; $\vec J$ is the current density) and in magnitude larger than the cylindrical-radial component of the centrifugal force density $F_{cR}$, that is, the regions where the Lorentz force tends to bring the flow closer to the rotation axis. The overlapping of the green and purple contours shows that the inwardly-directed Lorentz force constitutes an example of magnetic hoop stress.

Re-collimation in massive protostellar outflows has been observed in the form of jet knots, in e.g. IRAS 21078+5211 \citep{Moscadelli2021multi} and the radio jet HH 80-81 \citep{Rodriguezkamenetzky2017}. In those systems, non-thermal synchrotron emission originated in the knots is observed, where the particles are accelerated until they reach relativistic speeds. We find velocities of the order of $\sim 1000 \unit{km\,s^{-1}}$ in the jet material, however, the highest speed of the ejecta cannot be determined from our results because of the Alfvén limiter we set up in order to be able to avoid extremely small time steps. Even though we cannot fully recover the kinematics that produce the nonthermal emission observed in jet knots, the re-collimation mechanism we observe provides an explanation and a model for the production of the knots themselves. For a concrete example, the re-collimation point \emph{a} in Fig. \ref{x16_jc}, located at around $1000\unit{au}$ away from the massive protostar, and where such high speeds are seen in the simulation data, coincides reasonably well in position with the nonthermal emission lobe (JVLA observations at $5\unit{cm}$) observed in IRAS 21078+5211 \citep[][fig. 5]{Moscadelli2021multi}. The re-collimation of the outflow, which restricts the width of the cavity, is also visible in the shape of the radio lobes at scales of $\sim 10\,000\unit{au}$, taken from SO molecular transitions which trace external shocks of the flow (e.g., coming from the cavity wall; see \citealt{Moscadelli2021multi}).

Three-dimensional effects could potentially intervene in the production of jet knots and the propagation of the outflow. These effects should be studied with three-dimensional simulations of a disk-jet system, where the assumption of axisymmetry is relaxed. Material from fragments formed in the accretion disk could potentially be ejected along the jet cavity, producing jet knots or bow shocks (however, this has not been reported in three-dimensional simulations in the literature; see Sect. \ref{S: previous}). Additionally, episodic accretion might cause small precession in the jet axis, as suggested by the different regions of the red- and blueshift in the ejected material at large and small scales in IRAS 21078+5211 \citep{Moscadelli2021multi, Moscadelli2022}.

\subsection{Bow shock and the formation of the cavity} \label{S: bow}

\begin{figure}
	\centering
	\includegraphics[width=\columnwidth]{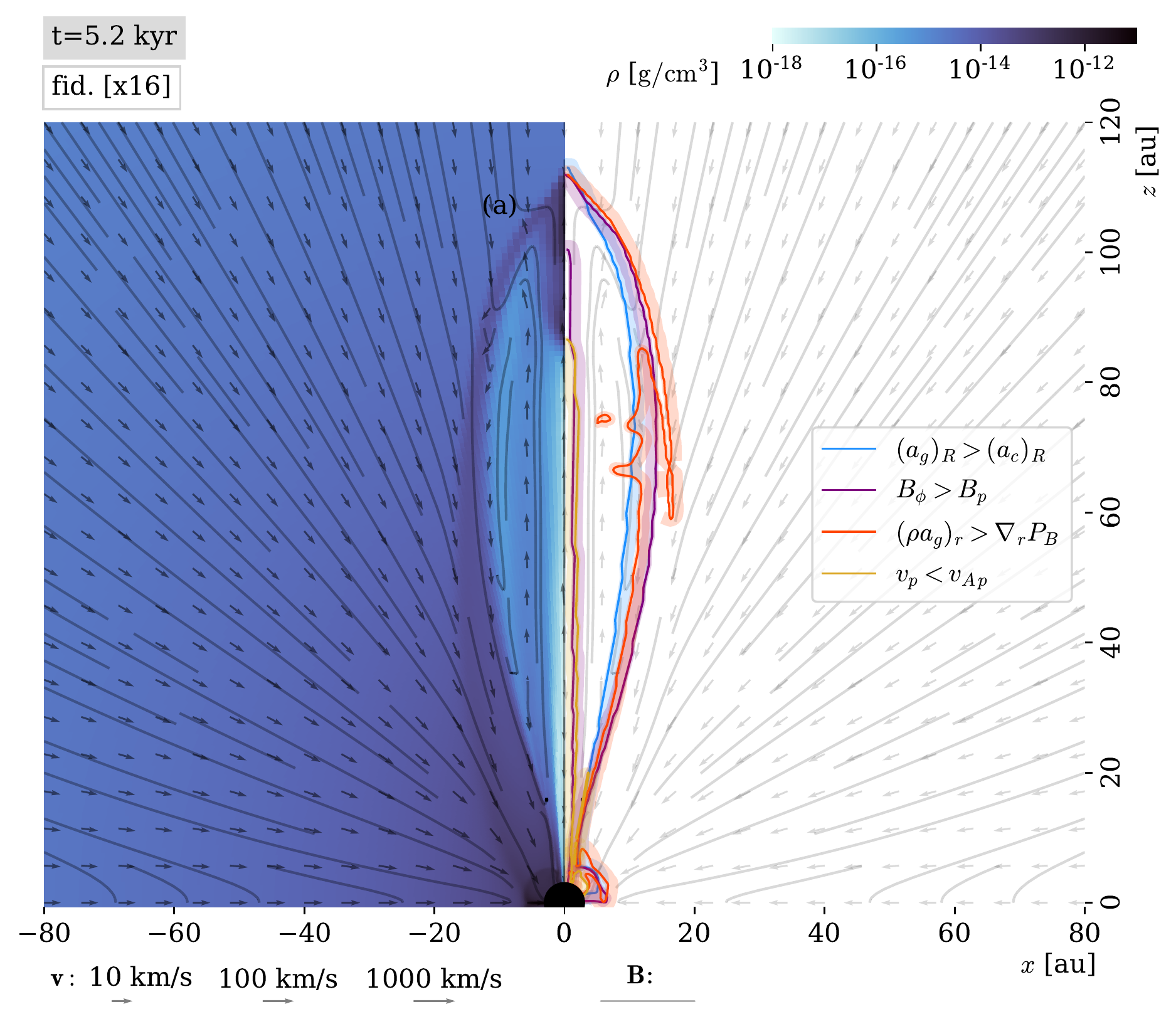}
	\caption{Production of the bow shock that gives rise to the jet cavity.}
	\label{x16_bow}
\end{figure}

Given that the magneto-centrifugal mechanism requires the existence of the cavity, the natural question to ask is how is this cavity formed in the first place. The formation of the cavity is captured in Fig. \ref{x16_bow}. The disk starts to appear shortly before the instant of time depicted (where it is only a few astronomical units in size). In the thick disk, the magnetic field is dragged by rotation and gravity, and magnetic pressure increases rapidly, until it eventually overcomes gravity and the outflow is launched. This is evidenced with the fact that the magnetic field in the cavity (excluding the region surrounding the rotation axis) is toroidal, and that the magnetic pressure gradient is larger in magnitude than gravity in the spherical radial direction. The velocities achieved in this stage are of the order of $10\unit{km\,s^{-1}}$, so they are in line with the values observed in the tower flow stage. The formation of the cavity produces a bow shock, marked in the figure with the label \emph{a}, which propagates outwards as the cavity expands over time. At $t\sim 5.4\unit{kyr}$ in the fiducial simulation for the x16 grid, the cavity is broad enough such that the magneto-centrifugal mechanism is able to start from the cavity wall, as described in Sect. \ref{S: mc}. Then, the velocities of the jet increase to more than $100\unit{km\,s^{-1}}$.

Bow shocks are often observed in protostellar outflows (e.g. \citealt{Moscadelli2021multi}; \citealt{2015AA...581A...4L}). Even though we expect bow shocks to be originated by multiple causes (for example, episodic accretion), the formation of the cavity provides a production mechanism for bow shocks that are highly symmetrical and appear distant in the ejected material. Moreover, the position and propagation velocity of the bow shock could serve as a crude estimate for the age of a given protostellar disk-jet system, as we suggested in \cite{Moscadelli2022}. In that particular case, we estimated from the fiducial simulation on grid x16 that the propagation of the bow shock is of about $3\,300 \unit{au\,kyr^{-1}}$, and with this, that when the protostar is $\approx 5.24 \unit{M_\odot}$, the bow shock has propagated at around $\sim 30\,000\unit{au}$. This coincides reasonably well with the observed position of the bow shock at $\approx 36\,000\unit{au}$ away from the forming massive star.

\subsection{Magnetic braking: effects on the outflows}\label{s: mb}
\begin{figure*}
	\sidecaption
	\includegraphics[height=7.2cm]{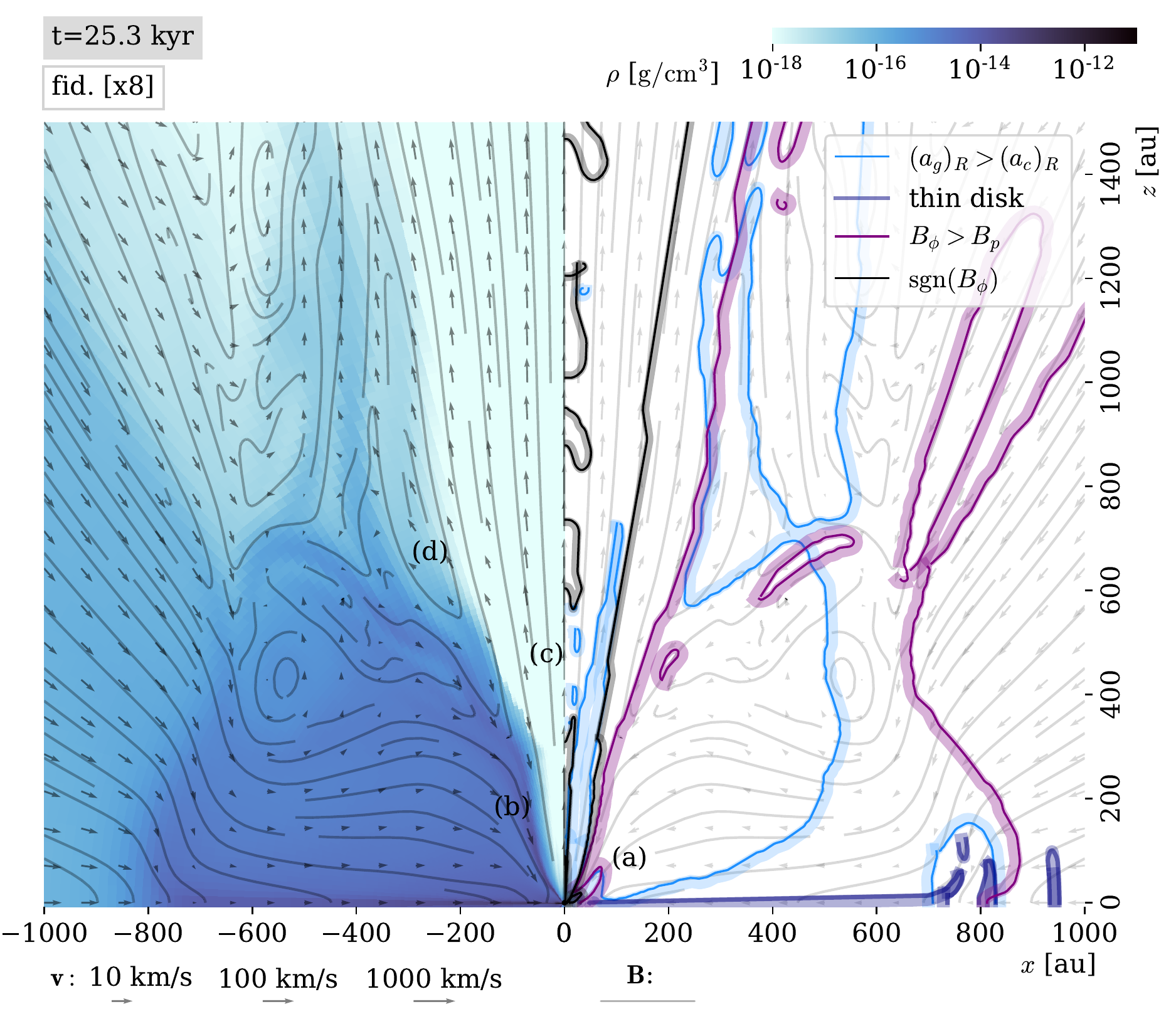}
	\includegraphics[height=7.2cm]{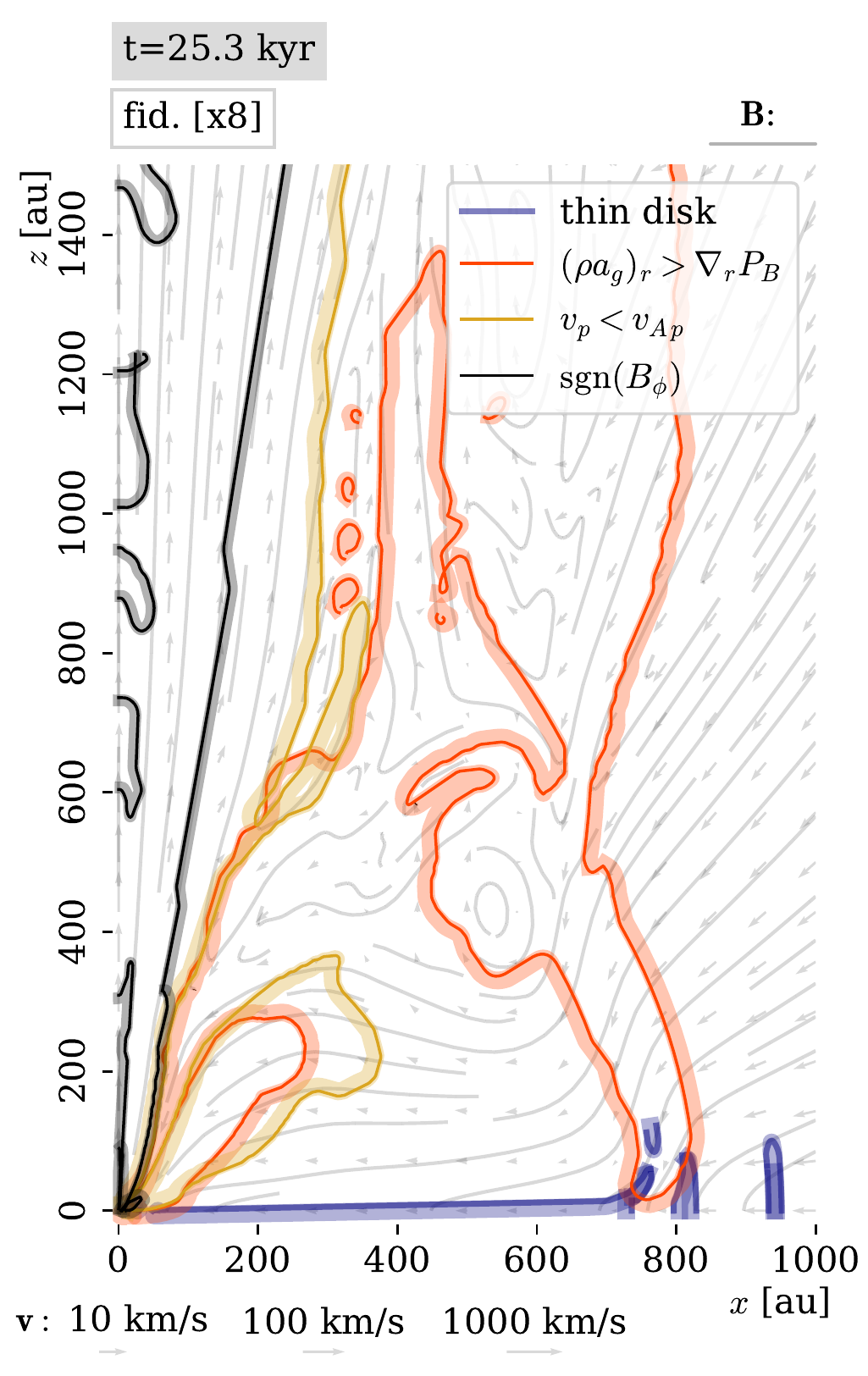}
	\caption{Effects of magnetic braking on the magnetically-driven outflows.}
	\label{x8_mb}
\end{figure*}

The rotational dragging of magnetic field lines creates magnetic tension, which tends to locally remove angular momentum very efficiently from the small scales in a process called magnetic braking \cite{Galli2006}. We discussed magnetic braking in the disk region in \citetalias{PaperI}, and showed that its effects are stronger as time passes and the magnetic field lines are progressively more wound. When enough angular momentum is removed from the innermost parts of the disk and cavity wall (which in our fiducial case happens at $t\sim 15\unit{kyr}$), those regions lose centrifugal support and the flow becomes mostly infalling. Figure \ref{x8_mb} shows an extensive morphological and dynamical analysis of the situation. The blue directed contour (gravito-centrifugal equilibrium) reveals the lack of centrifugal support for the inner parts of the disk (mark \emph{a}) and the cavity wall (mark \emph{b}), which corresponds to the magnetically braked flow region.

The magnetic field is mostly toroidal in the thin and thick layers of the disk, however, in the magnetically braked region of the disk it becomes mostly poloidal, following the flow. As time progresses, magnetic tension continues to transport angular momentum outwards from the braking radius of the disk until the full depletion of angular momentum in the braked region. When this happens, sections in the low-density cavity are observed to counter-rotate (with the corresponding sign reversal in the toroidal magnetic field; see black directed contour in Fig. \ref{x8_mb}). The counter-rotation is intermittent in the early stages. Counter-rotation in jets has previously been studied as a theoretical possibility \citep{Sauty2012, Cayatte2014, Staff2015}, and some observational evidence has been gathered (see, e.g. \citealt{Coffey2011}; \citealt{Louvet2016}) although it is not conclusive \citep{Tabone2020}.

The removal of angular momentum from the innermost scales interferes with the magneto-centrifugal mechanism as discussed in Sect. \ref{S: mc}. The center and right panels of Figure \ref{x8_mc} show that the lack of angular momentum in the inner parts of the thick disk and cavity wall (marker \emph{b}) impedes the magneto-centrifugal launching of material despite being in the sub-Alfvénic regime. As a consequence, the cavity becomes narrower close to the protostar. An outflow driven by magnetic pressure (marker \emph{c}) can still be launched from the innermost parts of the cavity (cf. orange contour, that shows dominance of magnetic pressure over gravity in the vertical direction). Despite the loss of angular momentum in the innermost scales, the magneto-centrifugal mechanism and cavity wall ejections are able to continue from higher points in the disk (marker \emph{d}), where the conditions of centrifugal support and sub-Alfvenicity are still satisfied. The tower flow, on the other hand, remains unaffected as the rotation of the disk continues to wind the magnetic field lines outside of the Alfvén surface, and therefore, magnetic pressure continues to be generated over time.

In this work, we did not include the effects of ambipolar diffusion and the Hall effect (see also the discussion in \S  \ref{S: param-B}). In particular, we expect ambipolar diffusion to have the following effects. Because ambipolar diffusivities increase with decreasing density \citep{Marchand2022} gravitational collapse would be promoted specially at its earliest stages. However, at the number densities found in the accretion disk ($10^9 \lesssim n \lesssim 10^{12} \unit{cm^{-3}}$), the diffusivity curves from \cite{Marchand2022} show that the ambipolar and Hall effect diffusivities behave similarly to Ohmic resistivity. Because of this, we would not expect a strong impact in the disk physics, although additional diffusivities might cause the thick layer of the disk to be thinner because of the decreased magnetic pressure. From \citetalias{PaperI}, we expect higher diffusivities to delay the effects of magnetic braking (see also \citealt{Masson2016} for an example study in the low-mass case where the interplay between ambipolar diffusion and magnetic braking is present).

Inside of the cavity, however, the situation is different. Even though ambipolar diffusivities increase with low densities and strong magnetic fields (such as inside the cavity), all magnetic diffusivities (Ohmic resistivity, ambipolar diffusivity and the Hall diffusivity) strongly drop when the material becomes ionized \citep{Marchand2022, Machida2007}. The ejected material from collimated outflows from massive (proto)stars has been found to be shock-ionized \citep{Moscadelli2021multi,Rodriguezkamenetzky2017}, so we expect our treatment of the jet physics in the cavity to be correct, especially at early times when magnetic braking effects are not yet present.

As the main focus of this article is the study of the mechanisms that mediate the production, propagation and termination of outflows, we point out that magnetic braking is nevertheless a mechanism capable of terminating a jet. Magnetic braking could then be the mechanism responsible for setting the timescale of collimated protostellar jets, and regulating the expansion of H\textsc{ii} regions, although this conclusion must be confirmed by the addition of more physical effects to the model.

\section{Variation of the outflow properties with initial conditions} \label{S: jet parameter scan}

\begin{figure*}
	\includegraphics[width=\textwidth]{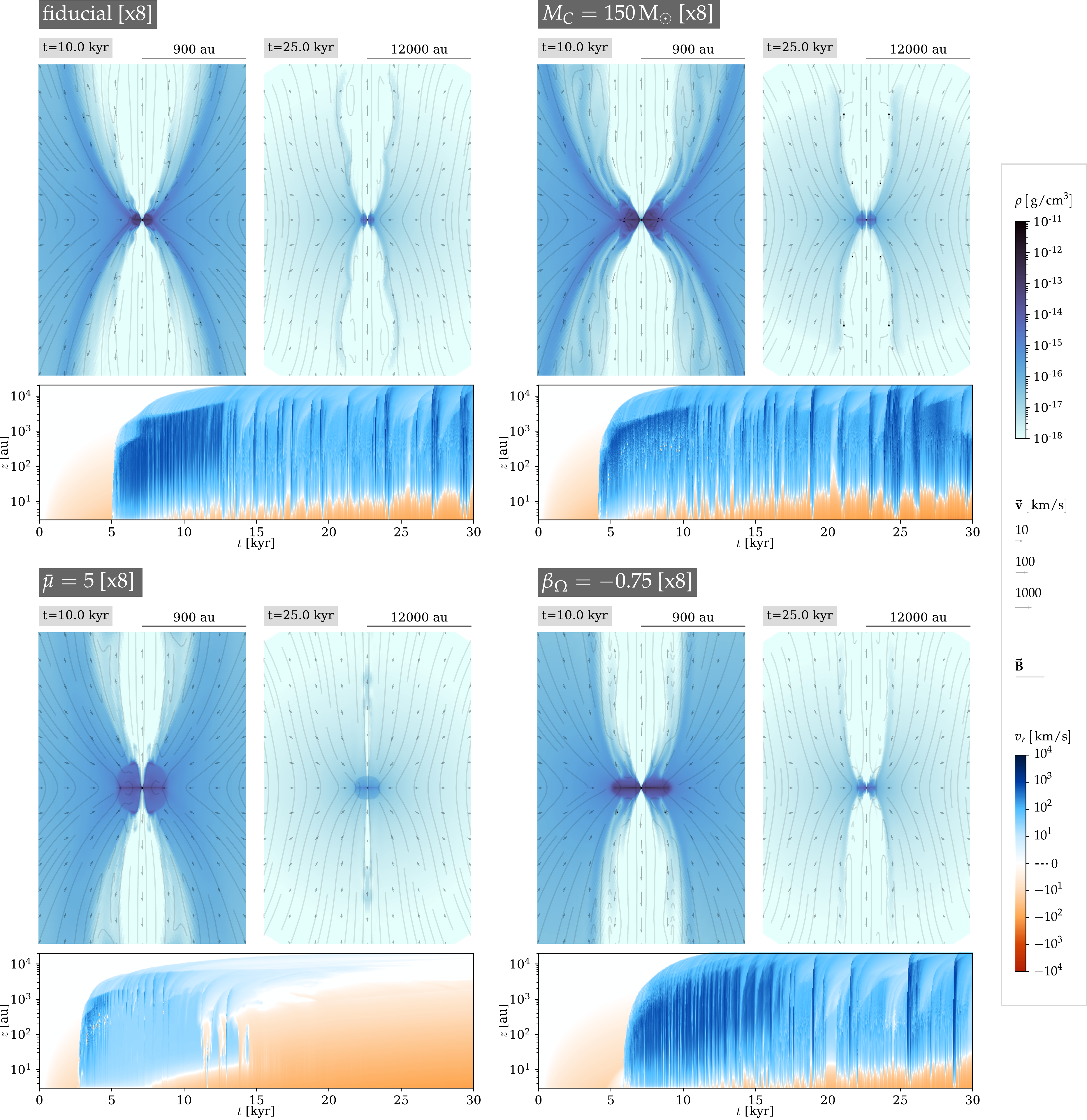}
	\caption{Morphology and kinematics of the outflows for different initial conditions, using the grid x8. The velocities shown as a function of time correspond to the maximum radial velocity within an angle of $22.5^\circ$ from the rotation axis, and the logarithmic scale has a cutoff of $\pm 1$ (velocities in the interval $-1\leq v_r\leq 1 \unit{km\,s^{-1}}$ are shown as zero).}
	\label{morph-x8}
\end{figure*}

\begin{figure*}
	\includegraphics[width=\textwidth]{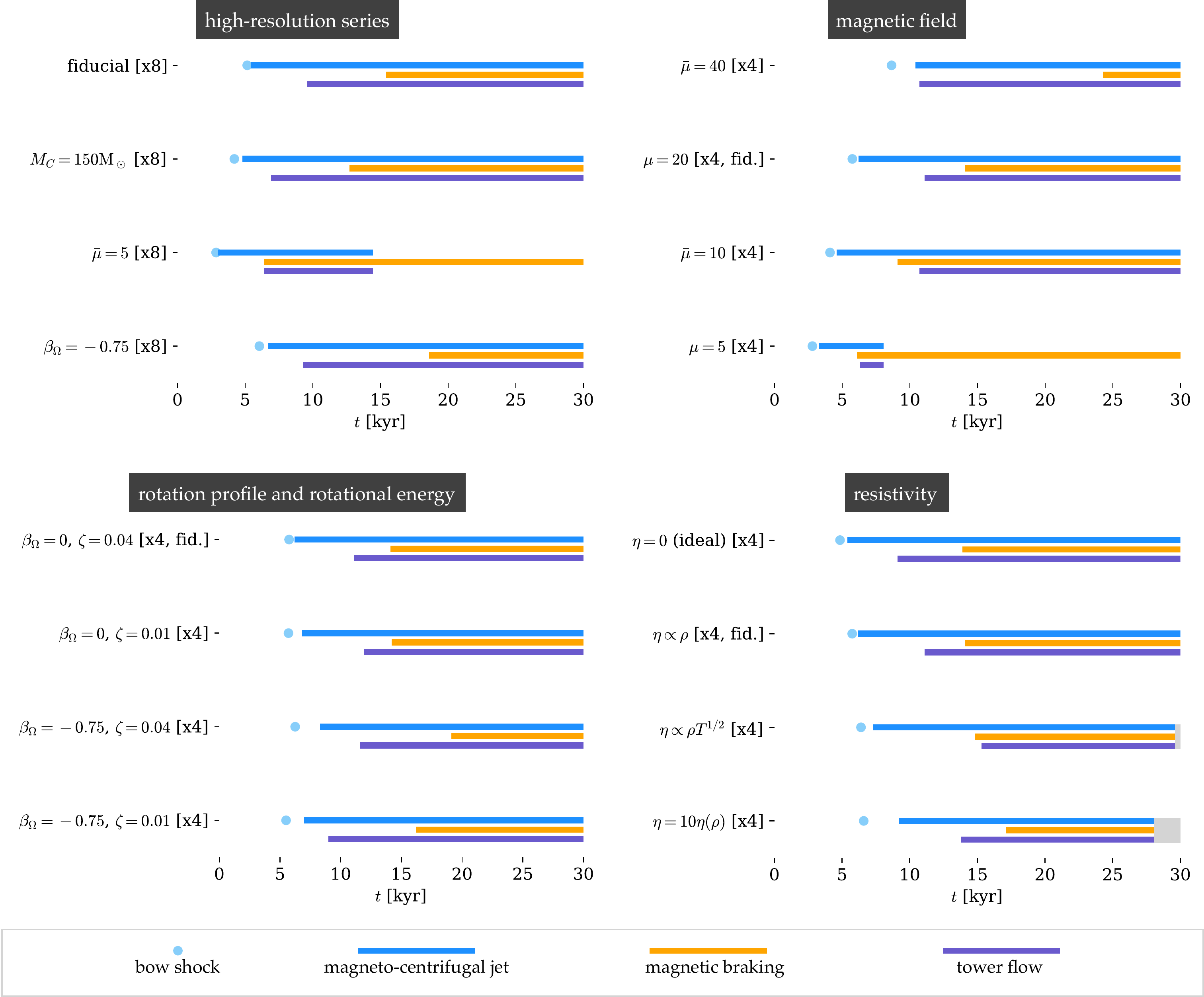}
	\caption{Comparison of the processes present in the jet for different initial magnetic field strengths, rotation profiles, rotational energies and resistivity models. For all the simulations except $M_C = 150\unit{M_C}\ \text{[x8]}$, the free-fall timescale is $52.4\unit{kyr}$. The gray boxes denote that the simulation was not run until $30\unit{kyr}$.}
	\label{processes}
\end{figure*}

\begin{figure*}
	\sidecaption
	\includegraphics[width=0.62\textwidth]{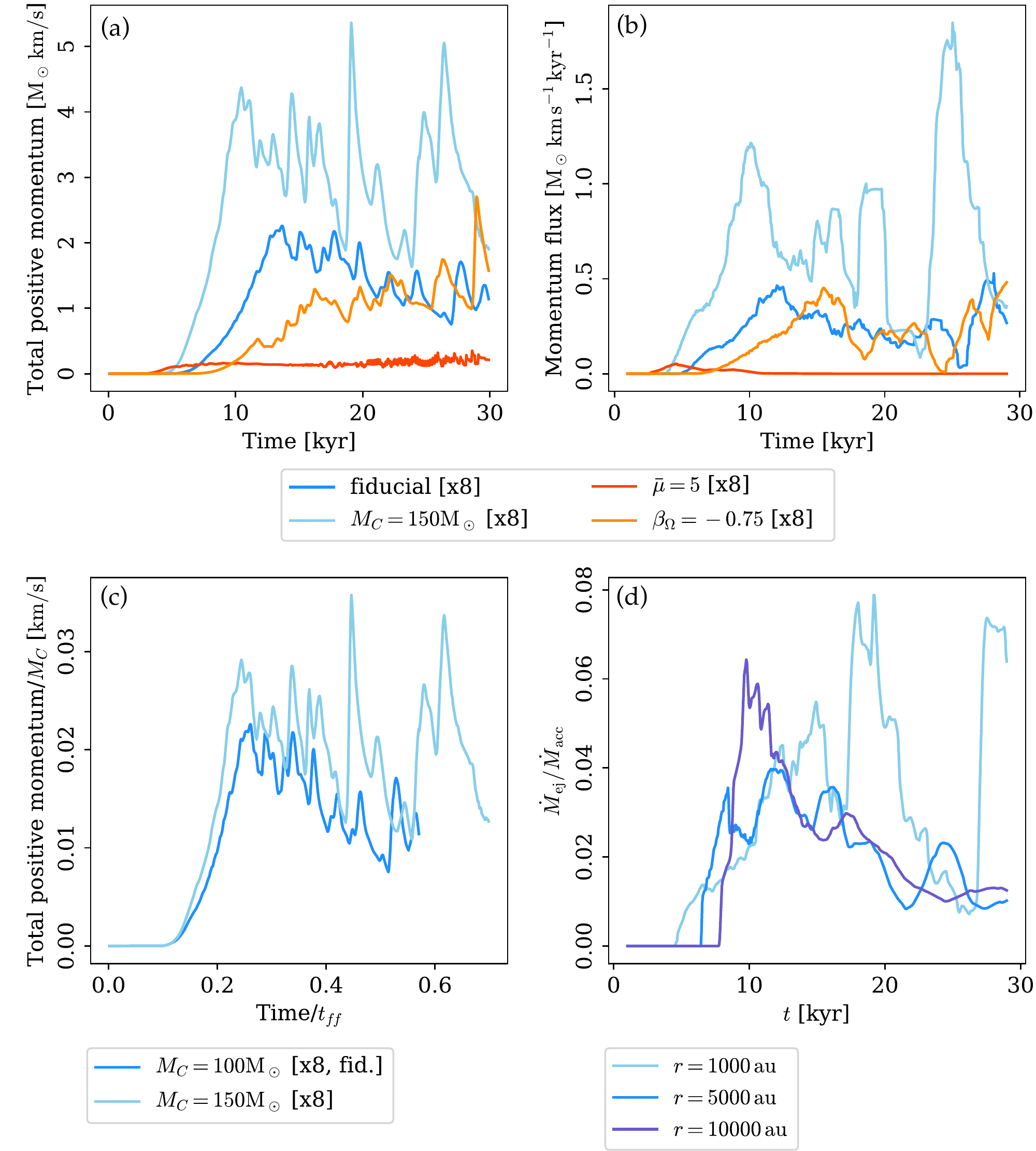}
	\caption{Panel \emph{a}: total momentum ejected by the magnetically-driven outflows within the cloud core, for the high-resolution series on grid x8. Panel \emph{b}: time-averaged momentum flux through a hemisphere of radius $r=15\,000\unit{au}$ for the same simulation series. Panel \emph{c}: total momentum ejected by the magnetic outflows, normalized to the mass of the cloud core, as a function of the fraction of the free-fall timescale elapsed, for two different masses of the cloud core. Panel \emph{d}: ejection-to-accretion rate measured using spheres of three different radii.}
	\label{jet-output}
\end{figure*}

In this section, we utilize the analysis from Sect. \ref{S: physics} and the results from other simulations in our series in order to investigate how the physical processes of the outflows are impacted by the conditions of the natal environment of the protostar. For comparisons between simulations, we use three kinds of graphics, which we present next.

Each panel of Fig. \ref{morph-x8} displays the morphology of the outflows at early and late times (with a corresponding adjustment to the spatial scale) accompanied by a velocity profile as a function of time. The velocity profile is taken along a cone of $22.5^\circ$ around the rotation axis, using the maximum value of the radial velocity at a given radial distance, and it permits us to study the kinematical signature of the jet and its propagation through the cloud core. In order to produce an effective morphological comparison, we used only the results for the high resolution grid (x8) because coarser grids in our series are not able to fully resolve all the scales of interest (see Appendix \ref{S: convergence}), although the basic physical processes are  still present.

Figure \ref{processes} exhibits a comprehensive overview of the temporal evolution of the dynamical processes discussed in Sect. \ref{S: physics}, namely, the formation of the cavity (leading to a bow shock), the magneto-centrifugal mechanism, magnetic braking and the magnetic tower flow. The time of launching of the bow shock is simply computed as when $v_r > 0$ for the first time at a reference altitude of $z=100\unit{au}$. The interval of presence of the magneto-centrifugal jet requires additionally that $(a_c)_R > (a_g)_R$ and $v_p < v_{A\,p}$ at the same altitude. Magnetic braking, in the other hand, is graphed by using the magnetic braking radius computed in \citetalias{PaperI}. Finally, the presence of the tower flow is fixed by simultaneously enforcing the conditions $|B_\phi| > |B_p|$, $\nabla_r P_B > (\rho a_g)_r$, and $v_r > 0$ at an altitude of $z=5\,000\unit{au}$.

Finally, Figure \ref{jet-output} aims to aid in the understanding of the effects of the jet in their environment with time, using only the data from the high-resolution series x8. We calculate the momentum flux (panel \emph{c}) of the jet through a hemisphere of radius $15\,000\unit{au}$, which gives an idea of the mechanical feedback that the magnetically-driven outflows impart to their natal environment at larger scales. In terms of mass, panel \emph{d} presents the accretion-to-ejection rate for the fiducial case, computed using the mass flux through spheres at different scales. The time average of the accretion-to-ejection rate at large scales in the cloud core remains within 10\%. Both quantities are characterized by a strong variability over time, due to the episodic ejection events (specially at late times) caused by the cavity wall ejections; for this reason, a time average is necessary to study their overall behavior. A smoother quantity is the total radially-outwards momentum contained in a hemisphere (i.e., the total momentum expelled by one branch of the jet), and it is presented in panels \emph{a} and \emph{c}.

\subsection{Mass of the cloud core}

Morphologically, the fiducial case ($M_C = 100\unit{M_\odot}$) and the higher mass cloud core ($M_C = 150\unit{M_\odot}$) are very similar, except that the cavity is ``pinched'' in the former case compared to the latter, which indicates that the jet re-collimation does not occur in the same place for both simulations. The velocity profile along the rotation axis shows different stages of evolution of the system in the fiducial case: the initial gravitational infall before $t\approx 5\unit{kyr}$, the formation of the cavity and the launching of the magneto-centrifugal jet (velocities of the order of $100\unit{km\,s^{-1}}$ after $t \approx 5\unit{kyr}$). The velocity profile does not show the acceleration of the jet on the smallest scales because it occurs radially outwards (Sect. \ref{S: acc}) and therefore it is not enclosed in the conical region used to compute the profile. However, the profile does capture the re-collimation of the jet at scales of $10^2\text{--}10^3\unit{au}$ (Sect. \ref{S: jc}), where the velocity reaches values of the order of $1000\unit{km\,s^{-1}}$.

All the dynamical mechanisms related to the outflows (Fig. \ref{processes}) are present in both simulations, although they start earlier in time for the more massive case. The same can be seen when examining the total linear momentum carried out by the outflows (Fig. \ref{jet-output}a), with the additional remark that the mechanical feedback from the magnetic outflows is stronger for more massive cloud cores. In \citetalias{PaperI}, we concluded that the disk dynamics scale with the mass of the cloud core, when examining the wider range of cloud core masses of $50\text{--}200\unit{M_\odot}$ using the lower resolution grid x4. Motivated by that result, we scaled the curves of the higher resolution results for the linear momentum by dividing by the mass of the cloud and expressed the time as a fraction of the free-fall timescale (Fig. \ref{jet-output}c). We find that the momentum scales well with the mass of the cloud, and so, despite the non-scalable nature of the thermodynamics and magnetic diffusivities of the system and its self-gravity, the results reported in this article can be reasonably used for cloud cores in a wider interval of masses, possibly including low-mass star formation.

Taking into account the respective free-fall timescales for the two clouds ($52.4\unit{kyr}$ for $M_C = 100\unit{M_\odot}$ and $42.8\unit{kyr}$ for $M_C = 150\unit{M_\odot}$), we corroborate the scalability of the outflow dynamics as follows. The bow shock and subsequent magneto-centrifugal jet start at $t=5.2\unit{kyr} = 0.1t_\mathrm{ff}$ for $M_C = 100\unit{M_\odot}$, which corresponds to $t=4.2\unit{kyr}$ for $M_C = 150\unit{M_\odot}$ and is in agreement with the respective simulation result. Similarly, the tower flow reaches the height of $z=5\,000\unit{au}$ at $t=0.18 t_\mathrm{ff}$ in the fiducial case and at $t=0.16 t_\mathrm{ff}$ in the more massive cloud core. We had checked the scalability of magnetic braking with the fraction of free-fall timescale in \citetalias{PaperI} already.

The curve for the outflow total linear momentum is explained as follows. At $t\sim 0.1 t_\mathrm{ff}$, the magneto-centrifugal jet is launched, and the total linear momentum increases as the ejecta propagates throughout the cloud core. The outflow reaches a distance of $0.1\unit{pc}$ at around $0.2t_\mathrm{ff}$. Once this happens, the total momentum reaches a maximum value, with variations caused by changes in the re-collimation zone and episodic ejection events (cavity wall ejections). We see from Fig. \ref{morph-x8} that the high velocities associated to re-collimation and jet launching become less frequent at late stages, because of the effects of magnetic braking. This, together with the slower tower flow (which broadens the cavity at late stages), cause the total linear momentum to decline slightly over time for $t\gtrsim 0.35t_\mathrm{ff}$.

\subsection{Magnetic field} \label{S: param-B}

Our parameter space for the magnetic field responds to two different assumptions on the environmental conditions for massive star formation. In low-mass star formation, supercriticality has been observed in prestellar cores not yet under gravitational collapse, with measured magnetic field strengths of the order of a few tens of microgauss \citep[see, e.g.,][]{2022Natur.601...49C, Crutcher2012}, which mean mass-to-flux ratios of the order of a few. In the case of massive star formation, the pre-collapse conditions of the magnetic field are not well-known, and because of this, we assume cases of lower and higher magnetization. The lower magnetization case assumes that the same order of magnitude of the magnetic field strength observed in low-mass star formation is also present in the massive case and it constitutes our fiducial case. The higher magnetization case assumes that a similar mass-to-flux ratio as observed in the low-mass star formation would be present in the high-mass scenario.

In what follows, we compare the outcome of the fiducial run (normalized mass-to-flux ratio $\bar \mu = 20$, which corresponds to an initial magnetic field of $68\unit{\mu G}$) and a simulation with a stronger initial magnetic field ($\bar \mu = 5$, corresponding to $0.27\unit{mG}$). Stronger magnetic fields cause stronger and earlier magnetic braking as already discussed in \citetalias{PaperI}, and the additional support from magnetic pressure allows for the formation of a larger accretion disk. The outflow cavity is narrower at early times (Fig. \ref{morph-x8}) with strong magnetic fields, which can be attributed to the additional magnetic pressure which makes the observed thick layer of the disk thicker, constraining the cavity close to the protostar.

In Sect. \ref{S: bow}, we presented our finding that the outflow cavity is initially formed by magnetic pressure, launching a bow shock in the process. A comparison of the points and bars that indicate the bow shock and the magneto-centrifugal phase in Fig. \ref{processes} confirms that, with more magnetic pressure, the cavity is able to form earlier in time, followed by the jet. However, the magneto-centrifugal jet is shorter-lived in the case with $\bar \mu = 5$ (cf. its velocity profile, Fig. \ref{morph-x8}). Magnetic braking for a cloud core with $\bar \mu = 5$ acts earlier in time compared to when $\bar \mu = 20$. The extraction of angular momentum in the inner disk becomes very strong after $t\gtrsim 15\unit{kyr}$, which impedes the magneto-centrifugal mechanism to continue. Additionally, the magnetic braking starts to act in the thick disk as well (but not in the thin disk, see \citetalias{PaperI}) turning the flow into sub-Keplerian and the tower flow cannot form because the magnetic pressure derived from the rotational dragging of magnetic field lines is not enough to overcome gravity. The magnetic pressure originated from the dragging of magnetic field lines by gravity, on the other hand, makes the thick disk thicker. We conclude from those results that very strong magnetic fields have the power to terminate magnetically-driven outflows. Because of the lower jet speeds, the momentum output (Fig. \ref{jet-output}a) of the outflows is considerably weaker in the case of stronger initial magnetic fields.

Because ambipolar diffusivities increase with magnetic field strength, we expect the the higher magnetization case to be the most affected by the noninclusion of ambipolar diffusion in the present study. However, we remark that we still see the formation of a disk-jet system, and that magnetic braking, dependent on the initial magnetic field strength, would set a timescale for the formation and duration of the magneto-centrifugal phase of the outflows. For example, for low-mass star formation (where the $\bar \mu$ is of the order of a few) the jet duration (as a fraction of the freefall time) could be shorter. The interplay between jet termination, magnetic braking and angular momentum transport should be examined in future studies.

\subsection{Angular momentum content}
We performed simulations of various ratios of rotational to gravitational energy ($\zeta = 0.01, 0.04$) and initial distributions of angular velocity $\Omega \propto R^{\beta_\Omega}$, where $\beta_\Omega = 0$ (solid body rotation) and $\beta_\Omega = \beta_\rho/2= -0.75$ (steep initial rotation profile, as we considered in our simulations of a fragmenting accretion disk in \citealt{Oliva2020}). The fiducial case is $\beta_\Omega = 0, \, \zeta = 0.04$. In the high-resolution series, we only investigated the case where the rotational-to-gravitational energy ratio was kept the same but the angular momentum is initially more concentrated in the center of the cloud (steep rotation profile instead of the solid body assumption). In \citetalias{PaperI}, we found that the accretion disk becomes larger thanks to the greater initial availability of angular momentum close to the protostar, and that the thick disk is flatter. As a result of a larger accretion disk for $\beta_\Omega = -0.75$, we find an initially wider tower flow (Fig. \ref{morph-x8}), which translates into a more cylindrical outflow in both small and large scales compared to solid body rotation.

Both the velocity profile in Fig. \ref{morph-x8} and the total linear momentum (Fig. \ref{jet-output}a) indicate that the propagation speed of the bow shock and the jet material is slower for the steep angular velocity profile compared to solid body rotation, although the re-collimation speeds are similar in both cases. From \citetalias{PaperI}, we know that the mass of the protostar does not increase the same way in both cases. For a steep initial angular momentum profile, the early formation of the accretion disk slows down the accretion rate onto the protostar. As an example, at $t=10\unit{kyr}$, $M_\star = 3.1\unit{M_\odot}$ for the fiducial case, but only $1.77\unit{M_\odot}$ for $\beta_\Omega = -0.75$. Even though the magnetic field lines are dragged more by rotation with more availability of angular momentum, and therefore magnetic pressure should increase, the lower radial mass flux onto the protostar translates into a reduced dragging of magnetic field lines by gravity, which means that magnetic pressure actually decreases and the propagation of the outflow becomes slower. This difference can also be seen in the formation of the accretion disk: while in the fiducial case the magnetic outflows and the disk are formed almost simultaneously, in the case with $\beta_\Omega = -0.75$, the jet is launched with a delay with respect to the formation of the accretion disk (cf. Fig. \ref{processes} and Fig. 8 of \citetalias{PaperI}). The differences in protostellar mass are also responsible for the delay in magnetic braking (see \citetalias{PaperI}).

Overall, the total outwards momentum reaches similar values for both cases. After examining the full parameter space including the results for the lower resolution series x4, we find that even though the initial angular momentum can cause differences in the launching and propagation of the jet, all the dynamical processes seen in the fiducial case are also seen in cloud cores with a lower ratio of rotational to gravitational energy.

\subsection{Ohmic resistivity} \label{S: param-eta}

As a final exploration, we look at the effect of resistivity in the dynamical processes present in the outflows (Fig. \ref{processes}). We found in \citetalias{PaperI} that the presence of resistivity delays magnetic braking compared to the ideal MHD case (the delay is more noticeable when comparing against the protostellar mass as opposed to time). Apart from our standard case, that considers resistivity only as a function of density, we included the full resistivity formula by \cite{Machida2007}, which also depends on temperature and can cause an increase of a factor of 10--100 compared to our standard case. Additionally, we included a simulation where Ohmic resistivity is artificially high by a factor of 10. These four simulations provide insight to what happens when the resistivity model is varied, i.e., when more magnetic diffusivity is added. From Fig. \ref{processes}, we see that the jet and the tower flow are launched slightly later for the case with higher resistivity. This is expected, because the jet launching mechanism depends on flux freezing inside of the Alfvén surface and the dragging of magnetic field lines by rotation. However, Fig. \ref{processes} shows the presence of all processes in all cases.

\section{Comparison to previous numerical studies} \label{S: previous}
In \citetalias{PaperI}, we presented a comparison of our results for the accretion disk dynamics with several studies in the literature. Here we continue the comparison of our results for the dynamics of the magnetic outflows against the results found in the same studies.

\cite{Anders2018} performed simulations of a forming massive star starting from the collapse of a cloud core with magnetic fields, but considering an isothermal equation of state (in contrast to our treatment of radiation transport with the flux-limited diffusion approximation). We base our fiducial case on the setup proposed in that study, except that they use a rotational to gravitational energy ratio of 2\% (our fiducial case is of 4\%). The study by \cite{Anders2018} was the first to be able to clearly distinguish a fast, magneto-centrifugally launched jet from the slower magnetic tower flow in the context of massive star formation. They found evidence of the magneto-centrifugal launching of the jet, however, they applied a series of analytical criteria from the classic bead-on-a-wire jet theory from \cite{BlandfordPayne1982} (for which the effects of the infalling envelope are missing) to identify the jet launching area both at early and late times (when it moves upwards). Our explicit identification of the Alfvén surface and the gravito-centrifugal contour (cf. Fig. \ref{x8_mc}) enabled us to identify the launching region more accurately for all times and determine the effects of the ram pressure of the infalling envelope on the geometry of jet launching, which cause departures from the classic theory.  \cite{Anders2018} performed a convergence study using grids with increasing resolution, but their highest resolution grid is equivalent to our x2 grid, which means that in this study we are able to resolve more structures in the plasma. The insufficient resolution of the outer regions in the cloud core in their convergence study meant that the authors were not able to find convergence in several quantities, including the jet duration and the momentum transported by the magnetic outflows. In contrast, we find outflow momenta that tend to converge for the two highest resolution grids (x8 and x16), and the minimum cell size required for the adequate resolution of the tower flow at large scales (see Appendix \ref{S: resolution}). In \cite{Anders2018}, the artificial mass generated by the Alfvén limiter grew to unrealistic limits for some simulations. The higher resolution grids used in our study and fine tuning of the Alfvén limiter allow us to keep the artificial mass generated by the limiter to negligible levels (cf. Appendix \ref{S: Alfven}).

\subsection{Studies with ideal MHD}

The study conducted by \cite{Banerjee2007} consisted in 3D ideal MHD simulations of a slowly-rotating Bonnor-Ebert sphere of $168\unit{M_\odot}$ on $1.62\unit{pc}$ a combination of cooling and radiation diffusion for the gas thermodynamics and using the code FLASH with an adaptive mesh refinement (AMR) grid. Despite only obtaining a sub-Keplerian disk (probably due to strong magnetic braking due to the lack of magnetic diffusivity), they observe early pseudo-disk-driven magnetic winds of typical velocities of $4\unit{km\,s^{-1}}$, whose launching is credited to the tower flow mechanism.

A setup more comparable to ours is used in \cite{Seifried2012} ($100\unit{M_\odot}$ within a radius of $0.125\unit{pc}$ and initial solid body rotation), where the effects of different mass-to-flux ratios was investigated. They find outflow velocities of $\sim 5\unit{km\,s^{-1}}$, which are both very collimated (weakly magnetized runs) and poorly collimated (strong magnetization). The authors point to the existence of a ``fast'' outflow component launched by the magneto-centrifugal mechanism, which is detected by a launching criterion based on the energy equation (cf. Eq. \ref{e:eneq-1}). However, the velocities of the fast outflow are still of the order of $10\unit{km\,s^{-1}}$, which we credit to the coarse grid used (3D AMR grid with a minimum cell size of $4.7\unit{au}$) and the unconstrained effects that magnetic braking causes when no magnetic diffusivity is considered. They also observe an outflow driven by magnetic pressure, although the authors do not call it a tower flow, even though the whole cavity is dominated by the toroidal component of the magnetic field. The ejection-to-accretion rates at large scales are somewhat larger than our findings (they find $\dot M_\mathrm{ej}/\dot  M_\mathrm{acc} \lesssim 0.4$).

The study of \cite{Myers2013} considers a denser cloud core ($300\unit{M_\odot}$ within a radius of $0.1\unit{pc}$) and $\bar \mu = 2$, but supersonic turbulence instead of rotation. They include a gray flux-limited diffusion approximation to radiation transport. Even though the authors report on the existence of magnetically-driven outflows of up to $40\unit{km\,s^{-1}}$, they do not discuss their origin. A similar approach was taken by \cite{Rosen2020}, but with masses of the cloud core of $150\unit{M_\odot}$. However, the coarse spacial grid used (minimum cell size of $20\unit{au}$) impedes them to resolve the launching mechanisms of the magnetic outflows, adopting a sub-grid outflow injection approach instead. This means that we cannot directly compare our results with theirs.

\subsection{Studies including Ohmic resistivity}
The three-dimensional studies by \cite{Matsushita2017} and \cite{Machida2020} use the same Ohmic resistivity model by \cite{Machida2007} as we do. However, while we use treatment of radiation transport based on the flux-limited diffusion approximation, both studies model the thermodynamics of the dust and gas with a barotropic equation of state. \cite{Matsushita2017} start from a Bonnor-Ebert sphere of several masses that range from $32$ to $1542\unit{M_\odot}$ and radius $0.28\unit{pc}$ which rotate as a solid body and that are threaded with an initially-uniform magnetic field determined by normalized mass-to-flux ratios of $\bar \mu = 2$, $5$ and $10$. The authors report on outflows of magnetic origin associated to the outer disk, but do not discuss their origin in detail. The speeds obtained in the outflows are of the order of $10\unit{km\,s^{-1}}$. \cite{Machida2020} start from a similar setup, but expand the parameter space of the cloud core masses (from $11$ to $525\unit{M_\odot}$) and normalized mass-to-flux ratios (from 2 to 20). Additionally, they introduce an enhancement factor to the density profile to promote gravitational collapse and their cloud core radius is $0.2\unit{pc}$. The authors classify their outflows into successful, delayed and failed. They observe protostellar outflows failing to evolve and collapse by the strong ram pressure when a massive initial cloud is weakly magnetized; a similar preliminary finding was reported in \cite{Matsushita2017} with the conclusion that only strongly magnetized cloud cores can drive protostellar outflows. This trend is opposite to our findings: we see that a relatively weak initial magnetic field ($\bar \mu = 20$) is sufficient to produce a magneto-centrifugally launched jet, while strong magnetic fields launch shorted-lived jets because of the increased action of magnetic braking over time. In \cite{Matsushita2017} and \cite{Machida2020}, the origin of the outflows is not thoroughly discussed, but given the magnitudes of the velocities obtained (a few tens of kilometers per second), it is reasonable to think that they are launched by magnetic pressure and not the magneto-centrifugal mechanism. The propagation of the failed outflows reported in \cite{Machida2020} are reported to be stopped by ram pressure from the infall, a direct consequence of the interplay between gravity and magnetic forces. We have seen both in this article and in \citetalias{PaperI} that the initial density and angular momentum distribution are crucial to determine the features of the accretion disk and the magnetically-driven outflows. For this reason, we speculate that the differences in the initial cloud mass distribution (they consider higher cloud masses which means stronger ram pressure) and the launching mechanisms for the outflows could be responsible for the contradictory conclusions obtained in our study and \cite{Matsushita2017} and \cite{Machida2020}.

\subsection{Studies including ambipolar diffusion}

Finally, we turn our attention to studies that include the effects of ambipolar diffusion but no Ohmic dissipation. \cite{Mignon-Risse2021II} found magnetically-driven outflows of velocities of the order of $30\unit{km\,s^{-1}}$ in simulations that start from a cloud core of $100\unit{M_\odot}$ within a radius of $0.2\unit{pc}$ (roughly equivalent to $M_C = 50$ in our setup), a density profile increasing towards the center of the cloud but with a plateau, solid body rotation, a rotational-to-gravitational energy ratio of 1\% and a mass-to-flux ratio of 5. Cloud cores with subsonic and supersonic turbulence were also considered in that study. The authors use the code RAMSES with an AMR grid of maximum resolution of $5\unit{au}$, including radiation transport using the flux-limited diffusion approximation and gray irradiation from the protostar(s). \cite{Commercon2022} start from a similar setup, but consider additional configurations with a rotational-to-gravitational energy ratio of 5\% and a normalized mass-to-flux ratio of 2. The characteristics of the outflows obtained in both studies correspond mostly to our wide tower flow rather than the fast jet. In \cite{Mignon-Risse2021II}, the authors determine that the origin of the outflows is indeed magnetic (as opposed to produced by irradiation from the protostar), and that they are driven by the magnetic pressure gradient. The authors argue that the magneto-centrifugal mechanism might be responsible for some of the high velocity flow present in their simulations, but do not perform an in-depth confirmation of this possibility. We show that very high resolution is needed close to the massive protostar in order to obtain a fast jet. On the other hand, their consideration of ambipolar diffusion might mean that the early stages of gravitational collapse and the late stages of the outflow dynamics are better represented in their simulations for low values of the mass-to-flux ratio.

\subsection{Sink particles and the launching region}
When comparing our results against studies that make use of sub-grid sink particle algorithms, such as the ones mentioned in this section with the exception of \cite{Anders2018}, it is important to take into account that those algorithms usually require the definition of an accretion radius. For example, the accretion radius in \cite{Commercon2022} is four times the minimum cell size, i.e., $\sim 20\unit{au}$. This means that the dynamics of the gas are only expected to be fully consistent with the system of equations beyond that radius. The use of a sink cell of $3\unit{au}$ in radius and a logarithmically spaced grid permitted us to resolve the launching region of the magneto-centrifugal jet in a more consistent way in the regions very close to the protostar. Additionally, the definition of an accretion radius when using sink particles affects any comparisons of the effects of magnetic braking, because we observe magnetic braking only in the innermost tens of astronomical units in the disk, within the size of the accretion radius of several of the studies above.

The minimum cell sizes of the simulations run in grids x8 and x16, of $0.06$ and $0.03\unit{au}$ respectively, have enabled us to directly compare our results to the VLBI observations of IRAS 21078+5211, which have a resolution of $0.05\unit{au}$ \citep{Moscadelli2022}. This level of detail, in both observations and simulations, had not been obtained so far. The recent advances in observational techniques for studying protostellar jets evidences the need for future numerical studies to accurately describe the processes in the innermost $\sim 100\unit{au}$ from the forming massive star.

\section{Summary and conclusions} \label{S: summary}
We have paved the path toward a more complete theory of magnetically-driven massive protostellar outflows by performing simulations that self-consistently generate them, by analyzing their dynamics and evolution over time, and by relating the conditions of the onset of gravitational collapse with their production and propagation into larger scales. We performed a set of 31 axisymmetric magnetohydrodynamical simulations, including Ohmic dissipation as a nonideal effect, and radiation transport of the thermal emission and absorption of the dust and gas. We summarize our findings as follows:

\begin{itemize}
	\item The outflow cavity is originally formed by the increase of magnetic pressure as a product of the dragging of magnetic field by the gravitational collapse and rotation. In the process, a bow shock is thrust outwards. Once the cavity is formed and the Alfvén velocity increases inside, the magneto-centrifugal mechanism starts.
	\item The fast ($\gtrsim 100\unit{km\,s^{-1}}$) and highly collimated component of the outflow is launched by the magneto-centrifugal mechanism. Ram pressure from the infalling envelope and the thick layer of the accretion disk constrain the geometry of the launching region into a narrower configuration compared to the predictions of classical jet theory.
	\item The jet material is accelerated along the magnetic field lines while its flow is sub-Alfvénic. Once it leaves the Alfvén surface, it enters a helical trajectory along the cavity.
	\item We observe episodic ejection events originating in the innermost parts of the cavity wall that are in contact with the thick layer of the accretion disk.
	\item Magnetic pressure originating from the rotational dragging of magnetic field lines by the accretion disk drives a slower ($\sim 10 \unit{km\,s^{-1}}$) magnetic tower flow that broadens over time.
	\item Magnetic hoop stress and the ram pressure from the infalling envelope re-collimate the jet in regions that are in principle compatible with jet knots observed in protostellar jets, where the material reaches relativistic velocities and non-thermal radiation is produced.
	\item Magnetic braking modifies the innermost regions of the disk and the jet, narrowing the cavity, and it is capable of terminating the magnetically-driven outflows.
	\item Our results indicate that the ejected momentum scales with the mass of the initial cloud core and its free-fall timescale.
	\item The morphology and dynamics of the magnetically-driven outflows affected by the initial conditions for the gravitational collapse. More centrally-concentrated angular momenta distributions, which produce a larger disk, drive a more cylindrical jet. Stronger initial magnetic fields launch jets produce narrower jets that last shorter periods of time.
\end{itemize}

By assuming axisymmetry and taking maximum advantage of a grid in spherical coordinates that is able to resolve the phenomena of interest in both large and small scales, we were able to obtain unprecedented detail in the inner $\sim 100\unit{au}$ around the protostar. This region is uncovered by a new generation of observational evidence \citep{Carrasco-Gonzalez2021, Moscadelli2021multi, Moscadelli2022}, which brings new challenges for theoretical predictions on massive protostellar outflows.

\begin{acknowledgements}
We thank Richard Nies for his contributions to the analysis of part of the dataset at the early stages of the project. AO acknowledges financial support from the Deutscher Akademischer Austauschdienst (DAAD), under the program Research Grants - Doctoral Projects in Germany, and complementary financial support for the completion of the Doctoral degree by the University of Costa Rica, as part of their scholarship program for postgraduate studies in foreign institutions. RK acknowledges financial support via the Emmy Noether and Heisenberg Research Grants funded by the German Research Foundation (DFG) under grant no.~KU 2849/3 and 2849/9.
\end{acknowledgements}

\bibliographystyle{aa} 
\bibliography{jetsfld}

\begin{thebibliography}{44}
\expandafter\ifx\csname natexlab\endcsname\relax\def\natexlab#1{#1}\fi

\bibitem[{{Banerjee} \& {Pudritz}(2007)}]{Banerjee2007}
{Banerjee}, R. \& {Pudritz}, R.~E. 2007, \apj, 660, 479

\bibitem[{{Beuther} \& {Shepherd}(2005)}]{Beuther2005}
{Beuther}, H. \& {Shepherd}, D. 2005, in Astrophysics and Space Science
  Library, Vol. 324, Astrophysics and Space Science Library, ed. M.~S.~N.
  {Kumar}, M.~{Tafalla}, \& P.~{Caselli}, 105

\bibitem[{{Beuther} {et~al.}(2020){Beuther}, {Soler}, {Linz}, {Henning},
  {Gieser}, {Kuiper}, {Vlemmings}, {Hennebelle}, {Feng}, {Smith}, \&
  {Ahmadi}}]{Beuther2020}
{Beuther}, H., {Soler}, J.~D., {Linz}, H., {et~al.} 2020, \apj, 904, 168

\bibitem[{{Blandford} \& {Payne}(1982)}]{BlandfordPayne1982}
{Blandford}, R.~D. \& {Payne}, D.~G. 1982, \mnras, 199, 883

\bibitem[{{Carrasco-Gonz{\'a}lez} {et~al.}(2010){Carrasco-Gonz{\'a}lez},
  {Rodr{\'\i}guez}, {Anglada}, {Mart{\'\i}}, {Torrelles}, \&
  {Osorio}}]{Carrasco-Gonzalez2010}
{Carrasco-Gonz{\'a}lez}, C., {Rodr{\'\i}guez}, L.~F., {Anglada}, G., {et~al.}
  2010, Science, 330, 1209

\bibitem[{{Carrasco-Gonz{\'a}lez} {et~al.}(2021){Carrasco-Gonz{\'a}lez},
  {Sanna}, {Rodr{\'\i}guez-Kamenetzky}, {Moscadelli}, {Hoare}, {Torrelles},
  {Galv{\'a}n-Madrid}, \& {Izquierdo}}]{Carrasco-Gonzalez2021}
{Carrasco-Gonz{\'a}lez}, C., {Sanna}, A., {Rodr{\'\i}guez-Kamenetzky}, A.,
  {et~al.} 2021, \apjl, 914, L1

\bibitem[{{Cayatte} {et~al.}(2014){Cayatte}, {Vlahakis}, {Matsakos}, {Lima},
  {Tsinganos}, \& {Sauty}}]{Cayatte2014}
{Cayatte}, V., {Vlahakis}, N., {Matsakos}, T., {et~al.} 2014, \apjl, 788, L19

\bibitem[{{Ching} {et~al.}(2022){Ching}, {Li}, {Heiles}, {Li}, {Qian}, {Yue},
  {Tang}, \& {Jiao}}]{2022Natur.601...49C}
{Ching}, T.~C., {Li}, D., {Heiles}, C., {et~al.} 2022, \nat, 601, 49

\bibitem[{{Coffey} {et~al.}(2011){Coffey}, {Bacciotti}, {Chrysostomou},
  {Nisini}, \& {Davis}}]{Coffey2011}
{Coffey}, D., {Bacciotti}, F., {Chrysostomou}, A., {Nisini}, B., \& {Davis}, C.
  2011, \aap, 526, A40

\bibitem[{{Commer{\c{c}}on} {et~al.}(2022){Commer{\c{c}}on}, {Gonz{\'a}lez},
  {Mignon-Risse}, {Hennebelle}, \& {Vaytet}}]{Commercon2022}
{Commer{\c{c}}on}, B., {Gonz{\'a}lez}, M., {Mignon-Risse}, R., {Hennebelle},
  P., \& {Vaytet}, N. 2022, \aap, 658, A52

\bibitem[{{Crutcher}(2012)}]{Crutcher2012}
{Crutcher}, R.~M. 2012, \araa, 50, 29

\bibitem[{{Galli} {et~al.}(2006){Galli}, {Lizano}, {Shu}, \&
  {Allen}}]{Galli2006}
{Galli}, D., {Lizano}, S., {Shu}, F.~H., \& {Allen}, A. 2006, \apj, 647, 374

\bibitem[{{Guzm{\'a}n} {et~al.}(2010){Guzm{\'a}n}, {Garay}, \&
  {Brooks}}]{Guzman2010}
{Guzm{\'a}n}, A.~E., {Garay}, G., \& {Brooks}, K.~J. 2010, \apj, 725, 734

\bibitem[{{K{\"o}lligan} \& {Kuiper}(2018)}]{Anders2018}
{K{\"o}lligan}, A. \& {Kuiper}, R. 2018, \aap, 620, A182

\bibitem[{{Kuiper} \& {Hosokawa}(2018)}]{KuiperHosokawa2018}
{Kuiper}, R. \& {Hosokawa}, T. 2018, \aap, 616, A101

\bibitem[{{Kuiper} {et~al.}(2010){Kuiper}, {Klahr}, {Beuther}, \&
  {Henning}}]{Kuiper2010circ}
{Kuiper}, R., {Klahr}, H., {Beuther}, H., \& {Henning}, T. 2010, \apj, 722,
  1556

\bibitem[{{Kuiper} {et~al.}(2011){Kuiper}, {Klahr}, {Beuther}, \&
  {Henning}}]{Kuiper2011}
{Kuiper}, R., {Klahr}, H., {Beuther}, H., \& {Henning}, T. 2011, \apj, 732, 20

\bibitem[{{Kuiper} {et~al.}(2020){Kuiper}, {Yorke}, \& {Mignone}}]{Kuiper2020}
{Kuiper}, R., {Yorke}, H.~W., \& {Mignone}, A. 2020, \apjs, 250, 13

\bibitem[{{Lefloch} {et~al.}(2015){Lefloch}, {Gusdorf}, {Codella},
  {Eisl{\"o}ffel}, {Neri}, {G{\'o}mez-Ruiz}, {G{\"u}sten}, {Leurini},
  {Risacher}, \& {Benedettini}}]{2015AA...581A...4L}
{Lefloch}, B., {Gusdorf}, A., {Codella}, C., {et~al.} 2015, \aap, 581, A4

\bibitem[{{Louvet} {et~al.}(2016){Louvet}, {Dougados}, {Cabrit}, {Hales},
  {Pinte}, {M{\'e}nard}, {Bacciotti}, {Coffey}, {Mardones}, {Bronfman}, \&
  {Gueth}}]{Louvet2016}
{Louvet}, F., {Dougados}, C., {Cabrit}, S., {et~al.} 2016, \aap, 596, A88

\bibitem[{{Lynden-Bell}(2003)}]{Lynden-Bell2003}
{Lynden-Bell}, D. 2003, \mnras, 341, 1360

\bibitem[{{Machida} \& {Hosokawa}(2020)}]{Machida2020}
{Machida}, M.~N. \& {Hosokawa}, T. 2020, \mnras, 499, 4490

\bibitem[{{Machida} {et~al.}(2007){Machida}, {Inutsuka}, \&
  {Matsumoto}}]{Machida2007}
{Machida}, M.~N., {Inutsuka}, S.-i., \& {Matsumoto}, T. 2007, \apj, 670, 1198

\bibitem[{{Marchand} {et~al.}(2022){Marchand}, {Guillet}, {Lebreuilly}, \& {Mac
  Low}}]{Marchand2022}
{Marchand}, P., {Guillet}, V., {Lebreuilly}, U., \& {Mac Low}, M.-M. 2022,
  arXiv e-prints, arXiv:2202.11625

\bibitem[{{Masson} {et~al.}(2016){Masson}, {Chabrier}, {Hennebelle}, {Vaytet},
  \& {Commer{\c{c}}on}}]{Masson2016}
{Masson}, J., {Chabrier}, G., {Hennebelle}, P., {Vaytet}, N., \&
  {Commer{\c{c}}on}, B. 2016, \aap, 587, A32

\bibitem[{{Matsushita} {et~al.}(2017){Matsushita}, {Machida}, {Sakurai}, \&
  {Hosokawa}}]{Matsushita2017}
{Matsushita}, Y., {Machida}, M.~N., {Sakurai}, Y., \& {Hosokawa}, T. 2017,
  \mnras, 470, 1026

\bibitem[{{McLeod} {et~al.}(2018){McLeod}, {Reiter}, {Kuiper}, {Klaassen}, \&
  {Evans}}]{McLeod2018}
{McLeod}, A.~F., {Reiter}, M., {Kuiper}, R., {Klaassen}, P.~D., \& {Evans},
  C.~J. 2018, \nat, 554, 334

\bibitem[{{Mignon-Risse} {et~al.}(2021){Mignon-Risse}, {Gonz{\'a}lez}, \&
  {Commer{\c{c}}on}}]{Mignon-Risse2021II}
{Mignon-Risse}, R., {Gonz{\'a}lez}, M., \& {Commer{\c{c}}on}, B. 2021, \aap,
  656, A85

\bibitem[{{Mignone} {et~al.}(2007){Mignone}, {Bodo}, {Massaglia}, {Matsakos},
  {Tesileanu}, {Zanni}, \& {Ferrari}}]{Mignone2007}
{Mignone}, A., {Bodo}, G., {Massaglia}, S., {et~al.} 2007, \apjs, 170, 228

\bibitem[{{Moscadelli} {et~al.}(2021){Moscadelli}, {Beuther}, {Ahmadi},
  {Gieser}, {Massi}, {Cesaroni}, {S{\'a}nchez-Monge}, {Bacciotti},
  {Beltr{\'a}n}, {Csengeri}, {Galv{\'a}n-Madrid}, {Henning}, {Klaassen},
  {Kuiper}, {Leurini}, {Longmore}, {Maud}, {M{\"o}ller}, {Palau}, {Peters},
  {Pudritz}, {Sanna}, {Semenov}, {Urquhart}, {Winters}, \&
  {Zinnecker}}]{Moscadelli2021multi}
{Moscadelli}, L., {Beuther}, H., {Ahmadi}, A., {et~al.} 2021, \aap, 647, A114

\bibitem[{{Moscadelli} {et~al.}(2022){Moscadelli}, {Sanna}, {Beuther}, {Oliva},
  \& {Kuiper}}]{Moscadelli2022}
{Moscadelli}, L., {Sanna}, A., {Beuther}, H., {Oliva}, A., \& {Kuiper}, R.
  2022, Nat. Astron., https://doi.org/10.1038/s41550-022-01754-4

\bibitem[{{Myers} {et~al.}(2013){Myers}, {McKee}, {Cunningham}, {Klein}, \&
  {Krumholz}}]{Myers2013}
{Myers}, A.~T., {McKee}, C.~F., {Cunningham}, A.~J., {Klein}, R.~I., \&
  {Krumholz}, M.~R. 2013, \apj, 766, 97

\bibitem[{{Obonyo} {et~al.}(2021){Obonyo}, {Lumsden}, {Hoare}, {Kurtz}, \&
  {Purser}}]{Obonyo2021}
{Obonyo}, W.~O., {Lumsden}, S.~L., {Hoare}, M.~G., {Kurtz}, S.~E., \& {Purser},
  S.~J.~D. 2021, \mnras, 501, 5197

\bibitem[{{Oliva} \& {Kuiper}(2020)}]{Oliva2020}
{Oliva}, G.~A. \& {Kuiper}, R. 2020, \aap, 644, A41

\bibitem[{{Oliva} \& {Kuiper}(2022)}]{PaperI}
{Oliva}, G.~A. \& {Kuiper}, R. 2022, \aap, submitted (Paper I)

\bibitem[{{Purser} {et~al.}(2021){Purser}, {Lumsden}, {Hoare}, \&
  {Kurtz}}]{Purser2021}
{Purser}, S.~J.~D., {Lumsden}, S.~L., {Hoare}, M.~G., \& {Kurtz}, S. 2021,
  \mnras, 504, 338

\bibitem[{{Purser} {et~al.}(2016){Purser}, {Lumsden}, {Hoare}, {Urquhart},
  {Cunningham}, {Purcell}, {Brooks}, {Garay}, {G{\'u}zman}, \&
  {Voronkov}}]{Purser2016}
{Purser}, S.~J.~D., {Lumsden}, S.~L., {Hoare}, M.~G., {et~al.} 2016, \mnras,
  460, 1039

\bibitem[{{Rodr{\'\i}guez-Kamenetzky}
  {et~al.}(2017){Rodr{\'\i}guez-Kamenetzky}, {Carrasco-Gonz{\'a}lez}, {Araudo},
  {Romero}, {Torrelles}, {Rodr{\'\i}guez}, {Anglada}, {Mart{\'\i}}, {Perucho},
  \& {Valotto}}]{Rodriguezkamenetzky2017}
{Rodr{\'\i}guez-Kamenetzky}, A., {Carrasco-Gonz{\'a}lez}, C., {Araudo}, A.,
  {et~al.} 2017, \apj, 851, 16

\bibitem[{{Rosen} \& {Krumholz}(2020)}]{Rosen2020}
{Rosen}, A.~L. \& {Krumholz}, M.~R. 2020, \aj, 160, 78

\bibitem[{{Sanna} {et~al.}(2019){Sanna}, {Moscadelli}, {Goddi}, {Beltr{\'a}n},
  {Brogan}, {Caratti o Garatti}, {Carrasco-Gonz{\'a}lez}, {Hunter}, {Massi}, \&
  {Padovani}}]{Sanna2019}
{Sanna}, A., {Moscadelli}, L., {Goddi}, C., {et~al.} 2019, \aap, 623, L3

\bibitem[{{Sauty} {et~al.}(2012){Sauty}, {Cayatte}, {Lima}, {Matsakos}, \&
  {Tsinganos}}]{Sauty2012}
{Sauty}, C., {Cayatte}, V., {Lima}, J.~J.~G., {Matsakos}, T., \& {Tsinganos},
  K. 2012, \apjl, 759, L1

\bibitem[{{Seifried} {et~al.}(2012){Seifried}, {Pudritz}, {Banerjee}, {Duffin},
  \& {Klessen}}]{Seifried2012}
{Seifried}, D., {Pudritz}, R.~E., {Banerjee}, R., {Duffin}, D., \& {Klessen},
  R.~S. 2012, \mnras, 422, 347

\bibitem[{{Staff} {et~al.}(2015){Staff}, {Koning}, {Ouyed}, {Thompson}, \&
  {Pudritz}}]{Staff2015}
{Staff}, J.~E., {Koning}, N., {Ouyed}, R., {Thompson}, A., \& {Pudritz}, R.~E.
  2015, \mnras, 446, 3975

\bibitem[{{Tabone} {et~al.}(2020){Tabone}, {Cabrit}, {Pineau des For{\^e}ts},
  {Ferreira}, {Gusdorf}, {Podio}, {Bianchi}, {Chapillon}, {Codella}, \&
  {Gueth}}]{Tabone2020}
{Tabone}, B., {Cabrit}, S., {Pineau des For{\^e}ts}, G., {et~al.} 2020, \aap,
  640, A82

\end{thebibliography}

\appendix

\section{Convergence of the results} \label{S: convergence}
A part of the aims of this study is to determine numerical parameter setups that produce consistent results for the disk and outflow dynamics. We present here a selection of the results in search for numerical convergence.

\subsection{Resolution} \label{S: resolution} 
\begin{figure*}
\includegraphics[width=\textwidth]{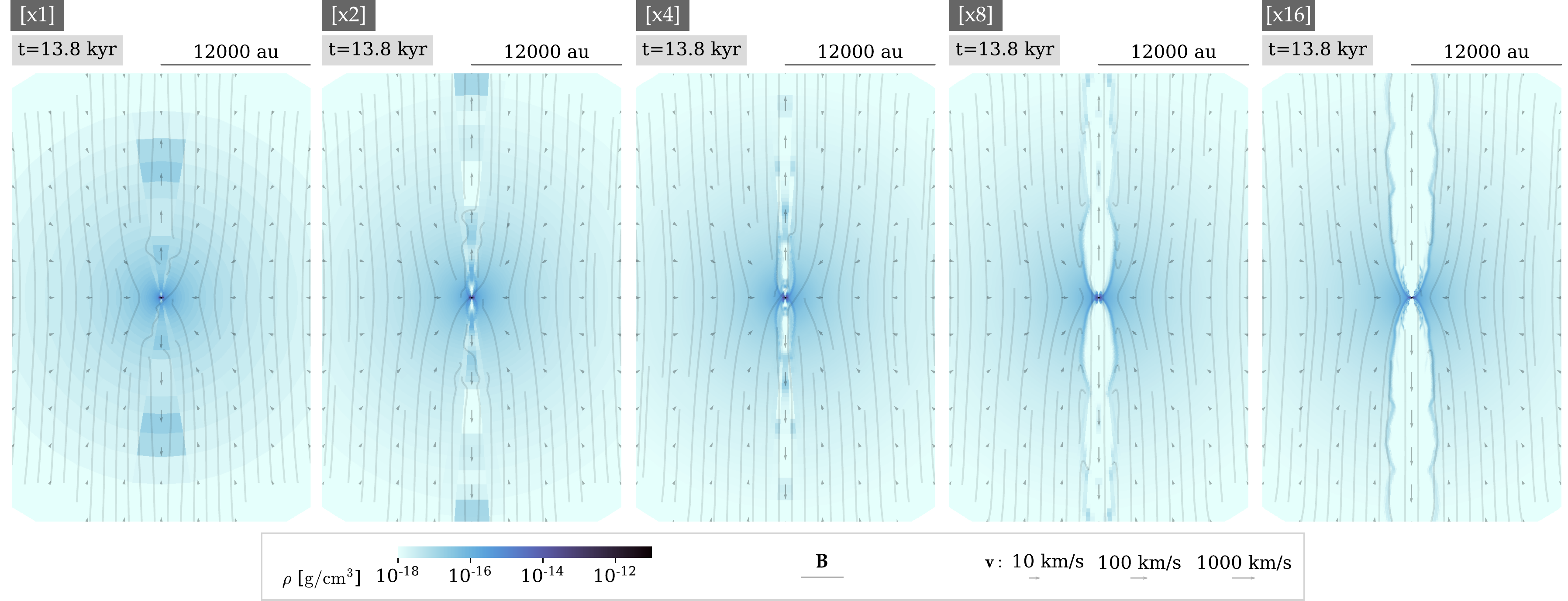}
\caption{Convergence of the jet cavity with resolution.}
\label{resol}
\end{figure*}

From a comparison between the results of \cite{Anders2018} and studies that use AMR Cartesian grids, it was clear that resolution of the jet launching region is extremely important for capturing the magneto-centrifugal mechanism and obtaining velocities of the order of $\sim 1000\unit{km\,s^{-1}}$. In the resolution study we performed, however, we observed several other critical phenomena to be properly resolved by spatial grid.

\begin{figure}
\includegraphics[width=\columnwidth]{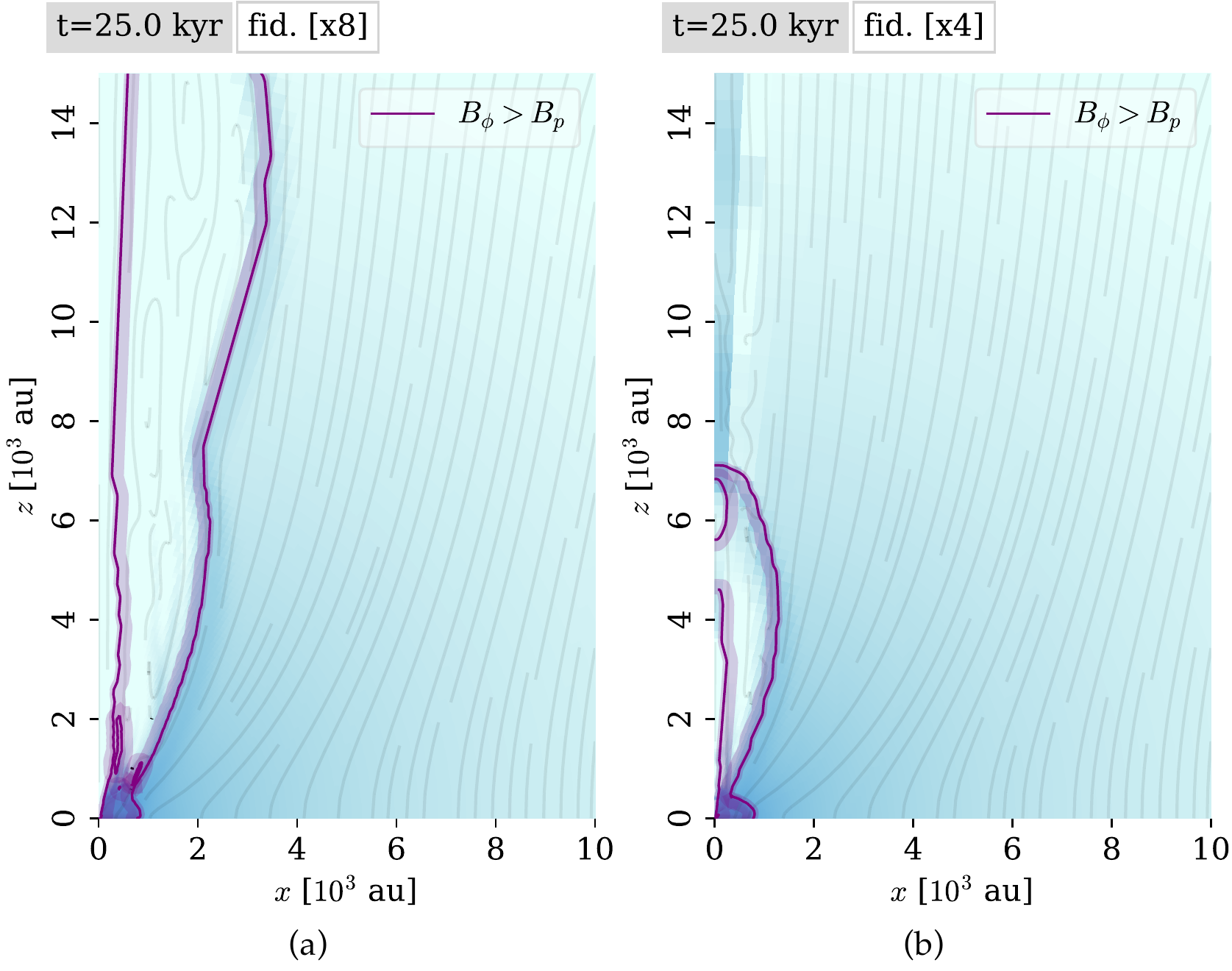}
\caption{Convergence of the tower flow with resolution at large scales.}
\label{resol-Bpt-cavity}
\end{figure}

A proper study of the propagation of the outflow is only possible if the large scales of the cavity are properly resolved. Figure \ref{resol} shows a comparison of the outflow cavity at large scales for all the grids in our investigation. Grid x16 produces an outflow cavity that is well resolved at large scales. The cavity contains the tower flow at large scales, and its thickness changes smoothly with distance. The results for grid x8 show most of the features of grid x16, although the thickness of the cavity does not change as smoothly as in simulation x16 for early times, even though it does for late times when the tower flow broadens. The outer parts of the cavity are resolved by $\sim 6$ grid cells. For the grid x4, geometry of the cavity changes qualitatively. In smaller scales, we observe the flow to be restricted to lobes, while in outer scales the cavity is narrower. A closer look to the outflow cavity for late stages (where the flow is dominated by the tower flow) is offered in Fig. \ref{resol-Bpt-cavity}. Beyond $z\sim 6\,000\unit{au}$, the cavity is not properly resolved (only 2-3 grid cells) and the toroidal nature of the magnetic field at large scales cannot be seen. The same is true for grids x2 (where the the lobes are even smaller) and x1 (where the propagation of the outflow itself is not properly represented any more.

\begin{figure}
\centering
\includegraphics[width=0.7\columnwidth]{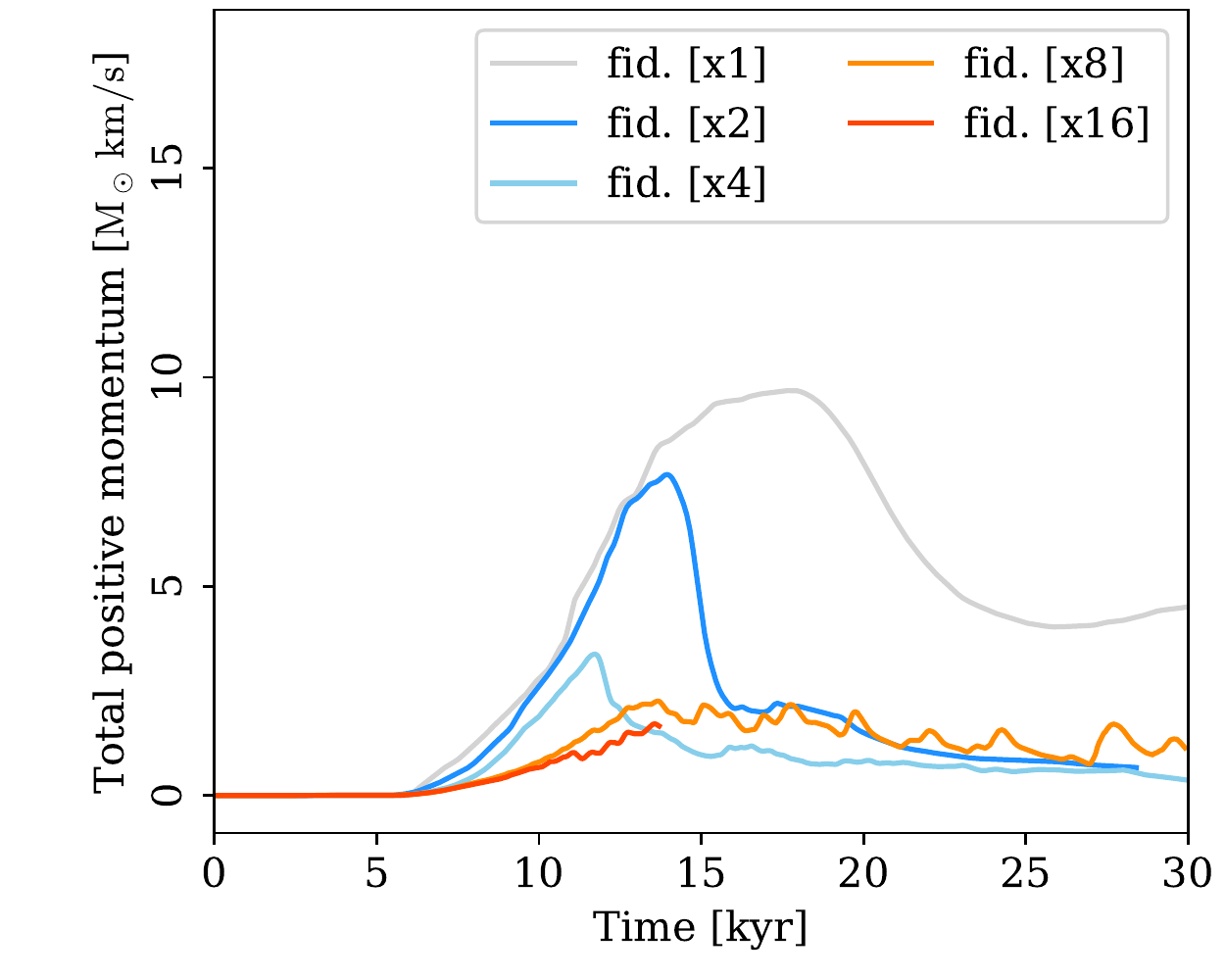}
\caption{Convergence of the total jet momentum with resolution.}
\label{resol_jetp}
\end{figure}

The total outflow momentum (Fig. \ref{resol_jetp}) is strongly hit by the resolution of the outflow cavity, but also by the minimum density resolved in the simulation. A higher resolution grid is able to produce lower densities as fine structures in the cavity are distinguished instead of being represented by their average density. Lower densities produce a weaker linear momentum. The curves for the total momentum in grids x4, x8 and x16 seem to agree in order of magnitude, which is attributable to a better agreement in the densities and velocities obtained. However, the curves also show a peak at early times only present in simulations x1, x2 and x4, which may be due to the improper resolution of the outflow cavity at large scales.

The narrowing of the cavity wall with the effects of magnetic braking was discussed in Sect. \ref{s: mb}. We note that, when this width is not properly resolved, the launching point of the magneto-centrifugal mechanism quickly moves outwards over time until the jet is completely terminated. This is the reason why in Fig. \ref{processes}, we only show the results for grids x4 and x8, where the cavity is partially and properly resolved at all times, respectively.

\subsection{Alfvén limiter} \label{S: Alfven}

As discussed in \citetalias{PaperI}, we use a varying density floor such that the Alfvén speed of a given grid cell does not increase beyond a fixed limit. This Alfvén limiter creates artificial mass as a result. We performed a parameter scan with different values of the maximum allowed Alfvén velocity with the following values: $v^A_\text{max}=200,500,1000,2000$ and $5000\unit{km\,s^{-1}}$. We monitored the total artificial mass generated in each case, and established a maximum acceptable value of $1\unit{M_\odot}$, i.e., 1\% of the mass of the cloud core, for the low-resolution exploratory runs. For grid x2, we observed that this limit was surpassed only for $v^A_\text{max}=200$ and $500\unit{km\,s^{-1}}$. For this reason, we chose $v^A_\text{max}=2000\unit{km\,s^{-1}}$ as the fiducial value for grids x1, x2 and x4. In grids x8 and x16, however, another parameter scan revealed that $v^A_\text{max}=1000\unit{km\,s^{-1}}$ yielded sufficiently low artificial masses (well below $0.005\unit{M_\odot}$), while saving computational costs, and therefore this value was chosen for the two highest-resolution runs. As a consequence, the artificial mass created by the Alfvén limiter has a negligible impact in the case of the high resolution runs we used for the main parameter scan in this study.

The Alfvén limiter is mostly active in the launching region of the jet (close to the massive protostar) and around the re-collimation regions. Therefore, we cannot conclude on the maximum velocities observed in a protostellar jet. This limitation of our setup, which is based on classical physics, leaves theoretical room for the relativistic velocities required for the nonthermal emission observed in re-collimation regions in protostellar jets (see, e.g., \citealt{Moscadelli2022} and \citealt{Rodriguezkamenetzky2017}).

\subsection{Sink cell size} \label{S: sink cell}
\begin{figure}
\includegraphics[width=\columnwidth]{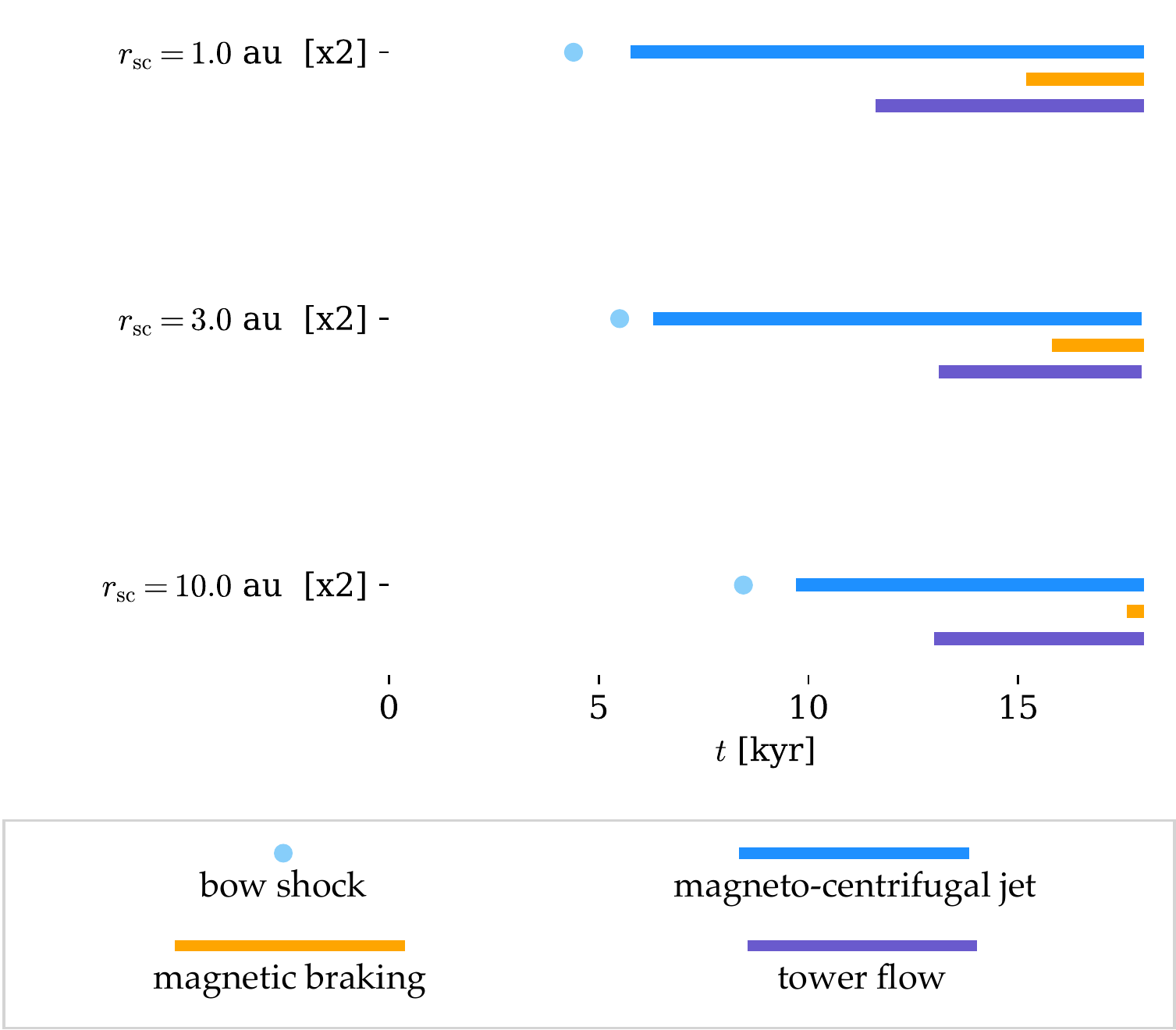}
\caption{Convergence of the observed dynamical processes of the magnetic outflows as a function of the size of the sink cell.}
\label{sc-processes}
\end{figure}

Finally, we wanted to check that our choice of a sink cell size does not fundamentally interfere with the physical processes described in this article. To this end, we performed a parameter scan with three values of the radius of the sink cell: $1\unit{au}$, $3\unit{au}$ (our fiducial case) and $10\unit{au}$. Due to computational costs, we only display the results for the first $15\unit{kyr}$ of evolution in Fig. \ref{sc-processes}. We find that the magneto-centrifugal launching of the jet and the formation of the bow shock occur earlier with a smaller sink cell. However, all the processes described in Sect. \ref{S: physics} are present in all the simulations, and the launching of the jet seems to happen still around $t\sim 5\unit{kyr}$ for the two smallest sink cell sizes.

\end{document}